\documentclass[reqno,11pt]{article}
\pdfoutput=1
\usepackage{jheppub}
\usepackage{epsfig}
\usepackage{amssymb}
\usepackage{verbatim}
\usepackage{amsmath}
\usepackage{hyperref}
\usepackage{dsfont}
\usepackage{slashed}
\usepackage[dvipsnames]{xcolor}
\usepackage{epstopdf}
\usepackage{graphicx}
\usepackage{graphicx}
\usepackage{caption}
\usepackage{subcaption}
\usepackage{mathtools}
\usepackage{stmaryrd}
\def\e{{\rm e}}

\def\d{\partial}

\newcommand{\be}{\begin{equation}}
\newcommand{\ee}{\end{equation}}
\newcommand{\bs}{\begin{split}}
\newcommand{\es}{\end{split}}
\newcommand{\ba}{\begin{align}}
\newcommand{\ea}{\end{align}}
\newcommand{\bea}{\begin{eqnarray}}
\newcommand{\eea}{\end{eqnarray}}
\newcommand{\bc}{\begin{center}}
\newcommand{\ec}{\end{center}}
\newcommand{\dd}{\mathrm{d}}

\title{Mellin-(Schwinger) representation of One-loop Witten diagrams in AdS}
\author[a,\,b]{Carlos Cardona}
\affiliation[a]{
Niels Bohr International Academy and Discovery Center, University of Copenhagen\\
The Niels Bohr Institute, Blegdamsvej 17, DK-2100, Copenhagen \O, Denmark\\
\&}

\affiliation[b]{Physics Division, National Center for Theoretical Sciences, National Tsing-Hua
University,\\ Hsinchu, Taiwan 30013, Republic of China.}
\emailAdd{carlosgiraldo@nbi.ku.dk}

\abstract{ In this paper we consider Witten diagrams at one loop in AdS space for scalar $\phi^3+\phi^4$ theory. After using Schwinger parametrization to trivialize the space-time loop integration, we extract the Mellin-Barnes representation for the one-loop corrections to the four-particle scattering up to an integration over the Schwinger parameters corresponding to the propagators of the internal particles running into the loop. We then discuss an approach to deal with those integrals.}

\begin{document}

\begin{flushright}
\vspace{10pt} \hfill{NCTS-TH/1713} \vspace{20mm}
\end{flushright}

\maketitle


\section{Introduction}\label{sect:intro}

Since the conception of the AdS/CFT correspondence \cite{Maldacena:1997re, Witten:1998qj}, the scattering of particles and strings on AdS spaces has become an important subject of study, not only because it would improve our understanding on scattering amplitudes in curved space-times, but also because it will teach us about the large-$N$ expansion of conformal correlation functions in theories with holographic interpretation. On those theories, an expansion of a given correlation function in terms of $1/N$ should corresponds to an expansion in loops Witten diagrams in AdS gravity. Therefore, the development of techniques to compute loop Witten diagrams is imperative for the study of large $N$ conformal field theories. In addition, very recently  \cite{Pasterski:2016qvg}, has been suggested that the on-shell kinematics  of the flat-space S-matrix can be described by momentum diagrams in AdS,   and in \cite{Cardona:2017keg} it was shown that indeed the factorization residues for the S-matrix in flat-space scalar theories might be described by Witten diagrams in AdS, implying that the perturbative flat-space S-matrix in $d+2$ dimensions might be related to correlation functions of some conformal field theory in $d-$dimension, which presents itself as a new interesting motivation to study Witten diagrams in AdS. 

 Right after the formulation of the AdS/CFT conjecture, some tree-level Witten diagrams were computed by conventional methods \cite{Arutyunov:2002fh,Freedman:1998bj,Liu:1998ty,Muck:1998rr,DHoker:1999kzh, Dolan:2000uw}, but more efficient and illuminating techniques were developed only recently starting with the introduction of the Mellin representation of correlation functions for conformal field theories constructed by Mack \cite{Mack:2009mi} and further developed in \cite{ Penedones:2010ue, Fitzpatrick:2011ia}. Even more recently, a new formalism called geodesic-Witten diagrams, that emulate the conformal block expansion of correlation functions from the gravity perspective, have been developed in \cite{Hijano:2015zsa} and extended to the inclusion of spinning particles in \cite{Dyer:2017zef,Sleight:2017fpc,Chen:2017yia, Castro:2017hpx, Nishida:2016vds}. However, despite those very interesting technical developments, the computation of loop Witten diagrams have proven to be very challenging and keeps being very unexplored. With the exception of few previous considerations \cite{Penedones:2010ue, Fitzpatrick:2011dm,Fitzpatrick:2011hu,Cornalba:2007zb} and some recent progress in, not even the most basic scalar loop integrals have been computed yet.

It is the purpose of this note to initiate a modest study on the computation of loop Witten diagrams from the gravity side of the duality. Through, to some extend, brute force approach, we intend to explicitly  pin point the basic obstacles preventing us to make progress in the perturbative computation of Witten diagrams and identify the building block integrals that we should focus in order to tackle the computation of loop Witten diagrams, hopefully in a wider scope that the ones considered here. 

The most recent studies of Witten diagrams in AdS are strongly dependent on the so-called split representation, on which every bulk-to-bulk propagator is, roughly speaking, replaced by an integration over a product of two bulk-to-boundary propagators, which at the loop level increase dramatically the number of integrations over propagators, converting a given loop diagram in an effective higher order loop diagram.
From the experience on flat-space loop computations, the integration over higher loops is, of course,  harder than the simplest one loop case, and therefore we don't want to rephrase a loop integral in terms of higher loops, but on the contrary, try to do the opposite if possible. 

In this note, we propose a different approach, by using directly bulk-to-bulk propagators in a representation that allow us to treat the bulk coordinates on AdS in almost the same footing as the boundary coordinates, avoiding us the introduction of additional  integrations over propagators. The advantage of this procedure is that loop-level diagrams are computed by essentially  following the same route used to compute tree-level diagrams. 

For concreteness, we focus on the simplest $\phi^3+\phi^4$ scalar theory on AdS. By using Schwinger parametrization to exponentiate the propagators we manage to integrate the space-time loop and subsequently extract the Mellin representation for the loop Witten diagrams corresponding to the corrections to the four particles scattering, expressed as function of Mellin variables plus a remaining integration over polynomials built out from Schwinger parameters associated to the internal propagators of the loop. Very recently, an algorithm to construct those polynomials at tree level and for the triangular loop,  has been developed in \cite{Nizami:2016jgt}. We identify those integrals as the hardest nut to crack on this approach and realized that better techniques should be developed to the evaluation of either, the particular integrals described in section 7 or the Witten diagram itself. We however press as hard as we can for the time being and for the case of the three particles scattering we manage to write the corresponding integral in terms of a lengthy combination of Hypergeometric functions. As well as, for the four particles scattering triangle and box, we sketch a proposal for the evaluation of the integrals by reinterpreting them as contour integrals. 

Recently the study of one loop Witten diagrams has been considered from the dual perspective by computing the $1/N$ expansion of scalar correlators in conformal field theories with holographic description  by means of the analityc bootstrap. More concretely, in \cite{Alday:2016njk} a method was developed to solve the crossing equations as an expansion in large spin. The given expansion can be resumed  and then extrapolated to lower spin (see also \cite{Caron-Huot:2017vep,Simmons-Duffin:2016wlq}).  This approach has been used in \cite{Alday:2017xua,Aharony:2016dwx,Aprile:2017bgs} to compute some one loop corrections to the four-point function of conformal fields with low conformal dimension, specifically $\Delta=2$ (for related recent work see also \cite{Li:2017lmh, Costa:2017twz}). It is our hope that we can reproduce those result in the near future by improving the approach initiated in this note.

After the completion of this work,  an interesting paper \cite{Giombi:2017hpr} appeared in the arXiv, where the authors compute explicitly the one loop correction to  2-point Witten diagrams.

The remainder of this paper is organized as follows. In section 2 we  illustrate the general approach by applying it to a simple tree-level scalar exchange. In section 3 we realized that the simplest loop correction, namely, the bubble Witten diagram follows from the scalar tree-level exchange without much effort. In section 4 we consider the one-loop correction to the three particle scattering, or in other words, to the structure constant for the scalar correlator. Section 5 and 6 deal with the most complicate diagrams for the loop correction to four particles scattering in $\phi^3+\phi^4$, namely, the four particle triangle  and the box. In section 7 we make a proposal to treat the leftover integral on loop-related Schwinger parameters.

\section{Warming up:  Tree-level scalar exchange }

 It is convenient to start by quickly introducing embedding coordinates \cite{Dirac:1936fq,Penedones:2007ns} since the computations quite simplify on those. Consider the embedding of euclidean AdS$_{d+1}$ in $(d+2)-$Minkowski space  $\mathbb{M}^{d+2}$, namely, $X$ lives in AdS$_{d+1}$ if for  $X\in\mathbb{M}^{d+2}$ it satisfies $X^2=-1$, where we have set the radius of AdS$_{d+1}$ space as $R=1$. The boundary of AdS$_{d+1}$ is given by null vectors $ P\in \mathbb{M}^{2+d}$, i.e $P^2=0$. 
The usual  parametrization of this coordinates is given by,
\be\label{embpara}
 X=(X^+,X^-,X^{\mu})={1\over z}(1,z^2+x^2,x^{\mu})\,,\quad P=(P^+,P^-,P^{\mu})=(1,x^2,x^{\mu})\,,
\ee 
where $x^{\mu}\in \mathbb{R}^d$.

\begin{figure}[h]
	\centering
  \includegraphics[width=7.0cm]{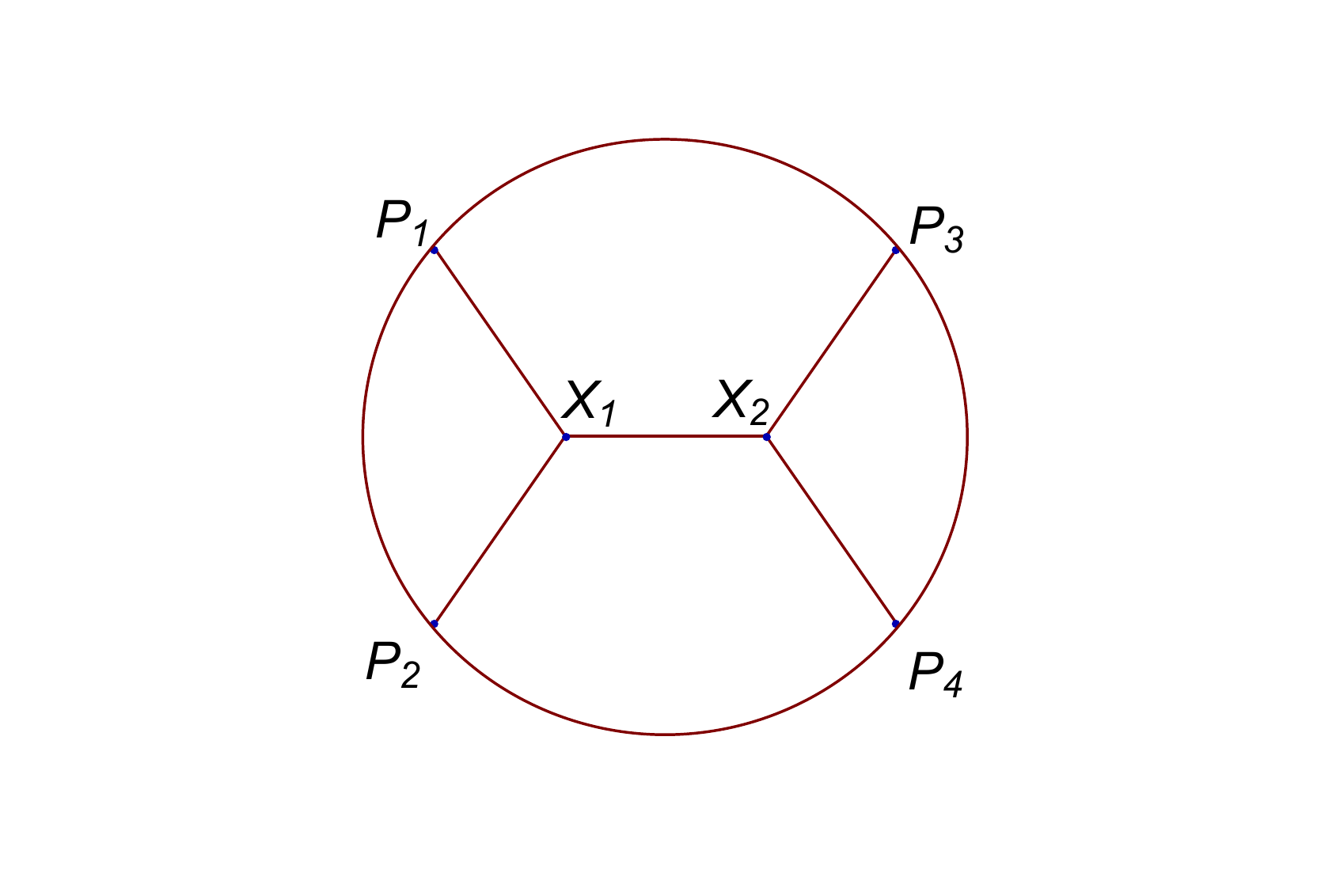}
  \caption{Scalar Exchange.}\label{fig01}
\end{figure}

Let us now consider the simple but still not-trivial Witten diagram shown in the figure \ref{fig01}.\footnote{The simplest not trivial diagram corresponds to a 4-point contact diagram.}, corresponding to a single scalar exchange.  We will use this example to illustrate the approach we are going to use in the following over more complicated examples.
This diagram was previously computed long time ago in \cite{DHoker:1999kzh} and later revisited in \cite{Penedones:2010ue} by means of the split-representation which we will review in the appendix A.
Let us recall some few basic facts about Witten diagrams, as the one represented in figure \ref{fig01}. The outer circle represents the boundary of $AdS$. Lines connecting points at the boundary to points in the bulk, represents bulk-to-boundary propagators which are given by
\be 
 G^{\Delta}_{\d B}(X, P)  = \frac{{\cal C}_\Delta}{(-2 P\cdot X)^\Delta} ,\quad{\cal C}_\Delta =\frac{\Gamma(\Delta)}{ 2\pi^h \Gamma(\Delta-h+1)}
\ee
where $\Delta$ denotes the conformal dimension of the conformal field at the boundary point $P$.
Lines connecting bulk points represents  bulk-to-bulk propagators, for which we have choosen the following representation \cite{Penedones:2010ue},
\be\label{normalbulk}
\begin{split}
 G^{\Delta}_{BB}(X_1, X_2)  = \int_{-i\infty}^{i\infty}{\dd c\,\dd \gamma\over (2\pi i)^2}{\Gamma(h+c-\gamma)\Gamma(h-c-\gamma)\over \Gamma(c)\Gamma(-c)((\Delta-h)^2-c^2)}{\Gamma(\gamma)\Gamma(h-\gamma)\over \Gamma(2h-2\gamma)\,u^{\gamma}}\,,
 \end{split}
\ee
where $u$ corresponds to the geodesic distance between the points $X_1$ and $X_2$ and is defined as,
\be
u\equiv(X_1-X_2)^2=-2(1+X_1\cdot X_2)\,. 
\ee
 Finally, bulk points have to be integrated over. Therefore, the amplitude associated to the diagram represented in figure \ref{fig01} is given by
\be\label{exchange}
\begin{split}
&{\cal A}^{\mathrlap{\Yright}~\Yleft}_4 (P_1, P_2, P_3, P_4) = \\
&g^2 \int_{AdS}\dd X_1 \dd X_2G^{\Delta_1}_{\d B}(P_1, X_1)  G^{\Delta_2}_{\d B}(P_2, X_1)  G^{\Delta}_{B B}(X_1, X_2)  G^{\Delta_3}_{B\d}(X_2, P_3) G^{\Delta_4}_{B\d}(X_2, P_4)\; .
\end{split}
\ee
where $g$ denotes the coupling of the particles entering the vertex of the diagram (see figure \ref{fig01}).
The crucial difference between the approach we are going to use in this note and previous computations (particularly the split representation approach) is that  we would like to treat the bulk coordinates $X$ at the same footing as the boundary coordinates $P$. In order to do that, we use the following Mellin-Barnes identity,
\be\label{mellinbarnes}
{1\over (A+B)^{\gamma}}={1\over 2\pi i \Gamma(\gamma)}\int_{-i\infty}^{i\infty}\dd f\, \Gamma(\gamma+f)\Gamma(-f){A^f\over B^{\gamma+f}}\,,
\ee 
to represent the bulk-to-bulk propagator as,
\be 
\begin{split}\label{mybulk}
 &G^{\Delta}_{BB}(X_1, X_2)  =\\
 & \int_{-i\infty}^{i\infty}{\dd c\,\dd \gamma \dd  f\over (2\pi i)^3}{\Gamma(h+c-\gamma)\Gamma(h-c-\gamma)\over \Gamma(c)\Gamma(-c)((\Delta-h)^2-c^2)}{\Gamma(\gamma+f)\Gamma(h-\gamma)\Gamma(-f)(-2)^f\over \Gamma(2h-2\gamma)\,(-2X_1\cdot X_2)^{\gamma+f}}\\
 &\equiv  \int_{-i\infty}^{i\infty}{\dd c\,\dd \gamma \dd  f\over (2\pi i)^3}\,\widehat{G}^{\Delta}_{BB}(c,\gamma,f)\,{1\over (-2X_1\cdot X_2)^{\gamma+f}}\,.
 \end{split}
 \ee
For the sake of notation we have defined,
 \be
 \widehat{G}^{\Delta}_{BB}(c,\gamma,f)= {\Gamma(h+c-\gamma)\Gamma(h-c-\gamma)\over \Gamma(c)\Gamma(-c)((\Delta-h)^2-c^2)}{\Gamma(\gamma+f)\Gamma(h-\gamma)\Gamma(-f)(-2)^f\over \Gamma(2h-2\gamma)}
 \ee
Putting this definitions back into \eqref{exchange}, we get
 \be\label{exchange2}
\begin{split}
&{\cal A}^{\mathrlap{\Yright}~\Yleft}_4 (P_i) =g^2 \prod_{i=1}^4{\cal C}_{\Delta_i} \int_{-i\infty}^{i\infty}{\dd c\,\dd \gamma \dd f\over (2\pi i)^3}\widehat{G}^{\Delta}_{BB}(c,\gamma, f)\\
& \int_{AdS}{\dd X_1 \dd X_2\over (-2P_1\cdot X_1)^{\Delta_1}  (-2P_2\cdot X_1)^{\Delta_2} (-2(X_1\cdot X_2))^{\gamma+f}  (-2P_3\cdot X_2)^{\Delta_3} (-2P_4\cdot X_2)^{\Delta_4}}\,.
\end{split}
\ee
The first observation from the representation \eqref{mybulk} is that it allows to think the amplitudes in the schematic form 
\be A\sim\int{\prod \dd X_i\over (X_j\cdot X_k)^{\Delta_{ij}}(X_r\cdot X_s)^{\Delta_{rs}}}\,,\ee
such us some of the $X_j's$ coordinates are projected to the boundary and the remaining should be integrated. We will see also that this representation is convenient when computing the integrals by using Schwinger parametrization.
The remaining of the computation proceed similarly as the computation of loop amplitudes in flat space-time. First, we introduce Schwinger parameters to exponentiate all the propagators as,
\be\label{exchange3}
\begin{split}
&{\cal A}^{\mathrlap{\Yright}~\Yleft}_4 (P_i) =g^2 \prod_{i=1}^4{\cal C}_{\Delta_i} \int_{-i\infty}^{i\infty}{\dd c\,\dd \gamma \dd f\over (2\pi i)^3}\widehat{G}^{\Delta}_{BB}(c,\gamma, f)\\
&\int^{\infty}_{0}\prod_{i=1}^5{\dd t_i\over t_i}{t_i^{\Delta_i}\over \Gamma(\Delta_i)} \int_{AdS}\dd X_1 \dd X_2\, \e^{2\left(\sum_{i=1}^2t_i\,P_i\cdot X_1+ t_5\,X_1\cdot X_2+\sum_{i=3}^4 t_i P_i\cdot X_2\right)}\,.
\end{split}
\ee
where for convenience of notation we have defined $\Delta_5={\gamma+f} $. Integration over bulk coordinates can be carried out straightforwardly now. All of them take the following form, 
\be \int \dd X_i \e^{2Q_i\cdot X_i}\ee
which has been computed in \cite{Penedones:2010ue}  and we review here. The key observation is that, since $Q$ is a vector in ${\mathbb M}^{d+2}$ and the integral should be Lorenz invariant, we can set $Q=|Q|(1,1,0)$. Then by using the parametrization \eqref{embpara} we can write,
\be\label{bulkintegral}
\begin{split}
\int \dd X_i \e^{2Q_i\cdot X_i}&=\int_0^{\infty}{\dd z_i\over z_i}z_i^{-d}\int \dd^d x_i\,\e^{(1+z_i^2+x_i^2)|Q_i|/z}\\
&=\pi^h\int_0^{\infty}{\dd z_i\over z_i}(z_i|Q_i|)^{-h}\e^{(1+z_i^2)|Q_i|/z_i}\\
&=\pi^h\int_0^{\infty}{\dd z_i\over z_i}z_i^{-h}\e^{-z_i+Q_i^2/z_i}\,,
\end{split}\ee
with $h\equiv d/2$.  Integration over $X_1$ have the form  \eqref{bulkintegral} with the corresponding $Q$ given by,
\be
Q_1=t_1 P_1+t_2 P_2 +t_5 X_2\,.
\ee
In order to factor out $Q_1^2$ from the  $z_1$ integration at the last line of \eqref{bulkintegral}, we rescalate $(t_1,t_2,t_5)\to\sqrt{z_1}(t_1,t_2,t_5)$ such as \footnote{Thought as part of the total integral \eqref{exchange3}.},
\be\begin{split}
 \int \dd X_1 \e^{2Q_1\cdot X_1}&=\pi^h\e^{Q_1^2}\int_0^{\infty}{\dd z_1\over z_1}\e^{-z_1}z_1^{\Delta_1+\Delta_2+\gamma+f-2h\over 2}\\
 &=\e^{Q_1^2}\Gamma\left({\Delta_1+\Delta_2+\gamma+f-2h\over 2}\right)\,.\end{split}\ee
 In the same manner, integration over $X_2$ produces an expression as  \eqref{bulkintegral} with 
 \be
 Q_2=t_1t_5 P_1+t_2 t_5 P_2+t_3 P_3+t_4 P_4\,, 
 \ee
 and again we re-escalate $(t_3,t_4,t_5)\to\sqrt{z_2}(t_3,t_4,t_5)$ to factor out the $Q_2$ dependence from the $z_2-$integral such as,
 \be\begin{split}
 \int \dd X_2 \e^{2Q_2\cdot X_2}&=\pi^h\e^{Q_2^2}\int_0^{\infty}{\dd z_2\over z_2}\e^{-z_2(1+t_5^2)}z_2^{\Delta_3+\Delta_4+\gamma+f-2h\over 2}\\
 &={\e^{Q_2^2}\over (1+t_5^2)^{\Delta_3+\Delta_4+\gamma+f-2h\over 2}}\Gamma\left({\Delta_3+\Delta_4+\gamma+f-2h\over 2}\right)\,.\end{split}\ee
Notice that the term $(1+t_5^2)$ comes from squaring $Q_2$, rescaling $z_2$ as above and the fact that $X_3^2=-1$ (or $X_i^2=-1$  for general bulk points). Later on we will see that those factors coming from  the square of the bulk coordinates are the ultimate responsible for the increasing difficulty in performing the integrals at loop level of Witten diagrams on AdS by the Schwinger parametrization approach. This is analogous to the situation in flat space, where theories containing only massless fields $P_i^2=0$ are much more simpler than massive theories, i.e containing  particles such as $X_i^2=-1$.

After integration over the bulk coordinates, we end up with
\be
\begin{split}
&{\cal A}^{\mathrlap{\Yright}~\Yleft}_4 (P_i) =g^2 \prod_{i=1}^4{\cal C}_{\Delta_i} \int_{-i\infty}^{i\infty}{\dd c\,\dd \gamma \dd f\over (2\pi i)^3}\widehat{G}^{\Delta}_{BB}(c,\gamma, f)\\
&\int^{\infty}_{0}\prod_{i=1}^5{\dd t_i\over t_i}{t_i^{\Delta_i}\over \Gamma(\Delta_i)} {1\over (1+t_5^2)^{\Delta_3+\Delta_4+\gamma+f-2h\over 2}}\Gamma\left({\Delta_1+\Delta_2+\gamma+f-2h\over 2}\right)\Gamma\left({\Delta_3+\Delta_4+\gamma+f-2h\over 2}\right)\\
&~~~~\e^{2(t_1t_2 t_5 P_1\cdot P_2+t_1 t_3 t_5 P_1\cdot P_3+t_2 t_3 t_5 P_2\cdot P_3+t_2 t_4 t_5 P_2 \cdot P_4+t_3 t_4 P_3\cdot P_4)}\,.
\end{split}
\ee
Now we can use the Symanzik formula  \cite{Symanzik:1972wj,Mack:2009mi},
\be\label{Syma}
\int_0^{\infty} \int^{\infty}_{0}\prod_{i=1}^n{\dd t_i\over t_i}t_i^{\Delta_i}\e^{\sum_{i<j}^nt_it_jQ_{ij}}={1\over 2(2\pi i)^{n(n-3)\over2}}\int_{\Sigma_n}\prod_{i<j}^n\Gamma(\delta_{ij})(Q_{ij})^{-\delta_{ij}}\,,
\ee
where $\Sigma_n$ is the ${n(n-3)\over2}-$dimensional Manifold  defined by the solution to the system,
\be\label{kinematics}
\delta_{ii}=-\Delta_i\,,\quad \sum_{i=1}^n \delta_{ij}=0\,.
\ee
Using this identity on the integration over the Schwinger parameters associated to the bulk-to-boundary propagators, namely $t_i$ with $i\in \{1,\cdots,4\}$, we get,
\be
\begin{split}
&{\cal A}^{\mathrlap{\Yright}~\Yleft}_4 (P_i) =g^2 \prod_{i=1}^4{\cal C}_{\Delta_i} \int_{-i\infty}^{i\infty}{\dd c\,\dd \gamma \dd f\over (2\pi i)^3}\widehat{G}^{\Delta}_{BB}(c,\gamma, f){1\over 2(2\pi i)^{2}}\int_{\Sigma_4}\prod_{i<j}^4{\Gamma(\delta_{ij})\over \Gamma(\Delta_i)}(P_i\cdot P_j)^{-\delta_{ij}}\\
&~~~~~~~~~~~~~\Gamma\left({\Delta_1+\Delta_2+\gamma+f-2h\over 2}\right)\Gamma\left({\Delta_3+\Delta_4+\gamma+f-2h\over 2}\right)\\
&~~~~~~~~~~~~~\int^{\infty}_{0}{\dd t_5\over t_5}{t_5^{\gamma+f-\delta_{13}-\delta_{14}-\delta_{23}-\delta_{24}}\over \Gamma(\gamma+f)}  (1+t_5^2)^{-\delta_{12}-{\Delta_3+\Delta_4+\gamma+f-2h\over 2}}
\end{split}
\ee
Integration over $t_5$ can be expressed as a beta function and results in,
\be
\begin{split}
&{\cal A}^{\mathrlap{\Yright}~\Yleft}_4 (P_i) =g^2 \prod_{i=1}^4{\cal C}_{\Delta_i} \int_{-i\infty}^{i\infty}{\dd c\,\dd \gamma \dd f\over (2\pi i)^3}\widehat{G}^{\Delta}_{BB}(c,\gamma, f){1\over 2(2\pi i)^{2}}\int_{\Sigma_4}\prod_{i<j}^4{\Gamma(\delta_{ij})\over \Gamma(\Delta_i)}(P_i\cdot P_j)^{-\delta_{ij}}\\
&~~~~~~~~~~\Gamma\left({\Delta_1+\Delta_2+\gamma+f-2h\over 2}\right)\Gamma\left({\Delta_3+\Delta_4+\gamma+f-2h\over 2}\right)\\
&~~~~~~~~~~
\frac{\Gamma \left(\frac{\gamma +f-\delta _{1,3}-\delta _{1,4}-\delta _{2,3}-\delta _{2,4}}{2} \right) \Gamma \left(\frac{-2 h+2 \delta
   _{1,2}+\delta _{1,3}+\delta _{1,4}+\delta _{2,3}+\delta _{2,4}+\Delta (3)+\Delta (4)}{2} \right)}{2 \Gamma \left(-h+\delta _{1,2}+\frac{\gamma
   +f+\Delta (3)+\Delta (4)}{2} \right)}\,,
\end{split} 
\ee
also the integration over $f$ can be performed straightforwardly leading to
\be\label{exchange4}
\begin{split}
&{\cal A}^{\mathrlap{\Yright}~\Yleft}_4 (P_i) ={1\over 2(2\pi i)^{2}}\int_{\Sigma_4}\prod_{i<j}^4{\Gamma(\delta_{ij})\over \Gamma(\Delta_i)}(P_i\cdot P_j)^{-\delta_{ij}}\\
&\times\,g^2 \prod_{i=1}^4{\cal C}_{\Delta_i} \int_{-i\infty}^{i\infty}{\dd c\,\dd \gamma\over (2\pi i)^2}{\Gamma(h+c-\gamma)\Gamma(h-c-\gamma)\over \Gamma(c)\Gamma(-c)((\Delta-h)^2-c^2)}{\Gamma(h-\gamma)\over \Gamma(2h-2\gamma)}\\
&~~\times\,
\frac{\Gamma \left(\frac{\gamma -s}{2}\right) \Gamma \left(\frac{-2 h+\gamma +\Delta _1+\Delta _2}{2}\right) \Gamma \left(\frac{-2 h+\gamma
   +\Delta _3+\Delta _4}{2} \right) }{ \Gamma \left(\frac{-s+\Delta
   _1+\Delta _2}{2} \right) \Gamma \left(\frac{-s+\Delta _3+\Delta _4}{2}\right)}\,.
\end{split}
\ee
Where $s$ and $t$ are defined by the following solutions of the system \eqref{kinematics} for $n=4$,
\be 
\begin{split}
\delta _{12}&= \frac{1}{2} \left(\Delta _1+\Delta _2-s\right)\,,\quad
\delta _{13}= \frac{1}{2} \left(\Delta _1+\Delta _3-t\right)\,,\\
\delta _{14}&= \frac{1}{2}
   \left(-\Delta _2-\Delta _3+s+t\right)\,,\quad
   \delta _{23}= \frac{1}{2} \left(-\Delta _1-\Delta _4+s+t\right)\,,\\
   \delta _{24}&= \frac{1}{2} \left(\Delta _2+\Delta
   _4-t\right)\,,\quad
   \delta _{34}= \frac{1}{2} \left(\Delta _3+\Delta _4-s\right)\,.
\end{split}
\ee

Following Mack \cite{Mack:2009mi} and knowing that Witten diagrams correspond to correlation functions on the boundary (or pieces of the whole correlator), we expect the following Mellin representation,
\be
A(P_i)={{\cal N}\over (2\pi i)^{n(n-3)\over 2}} \int_{\Sigma_n}\dd \delta_{ij}M(\delta_{ij})\prod_{i<j}^n\Gamma(\delta_{ij})(P_i\cdot P_j)^{-\delta_{ij}}\,,
\ee
with 
\be
{\cal N}={1\over 2}\Gamma\left({\sum_{i=1}^n\Delta_i-2h\over 2}\right)\prod_{i=1}^n {{\cal C}_{\Delta_i} \over\Gamma(\Delta_i)}
\ee
therefore comparing with \eqref{exchange4} we conclude that the Mellin amplitude for the exchaged scalar diagram can be represented by \footnote{In this case the Mellin amplitude does not depend on $t$ because we are considering only the $s-$channel exchange.},
\be
\begin{split}
&M^{\mathrlap{\Yright}~\Yleft}=g^2 \int_{-i\infty}^{i\infty}{\dd c\,\dd \gamma\over (2\pi i)^2}
{\Gamma(h+c-\gamma)\Gamma(h-c-\gamma)\over \Gamma(c)\Gamma(-c)((\Delta-h)^2-c^2)}{\Gamma(h-\gamma)\over \Gamma(2h-2\gamma)}\\
&~~
\frac{\Gamma \left(\frac{\gamma -s}{2}\right) \Gamma \left(\frac{-2 h+\gamma +\Delta _1+\Delta _2}{2}\right) \Gamma \left(\frac{-2 h+\gamma
   +\Delta _3+\Delta _4}{2} \right) }{\Gamma \left(\frac{-2 h+\Delta _1+\Delta _2+\Delta _3+\Delta _4}{2} \right) \Gamma \left(\frac{-s+\Delta
   _1+\Delta _2}{2} \right) \Gamma \left(\frac{-s+\Delta _3+\Delta _4}{2}\right)}\,.
\end{split}
\ee
\section{Bubble Witten diagram}
\begin{figure}[h]
	\centering
  \includegraphics[width=8.0cm]{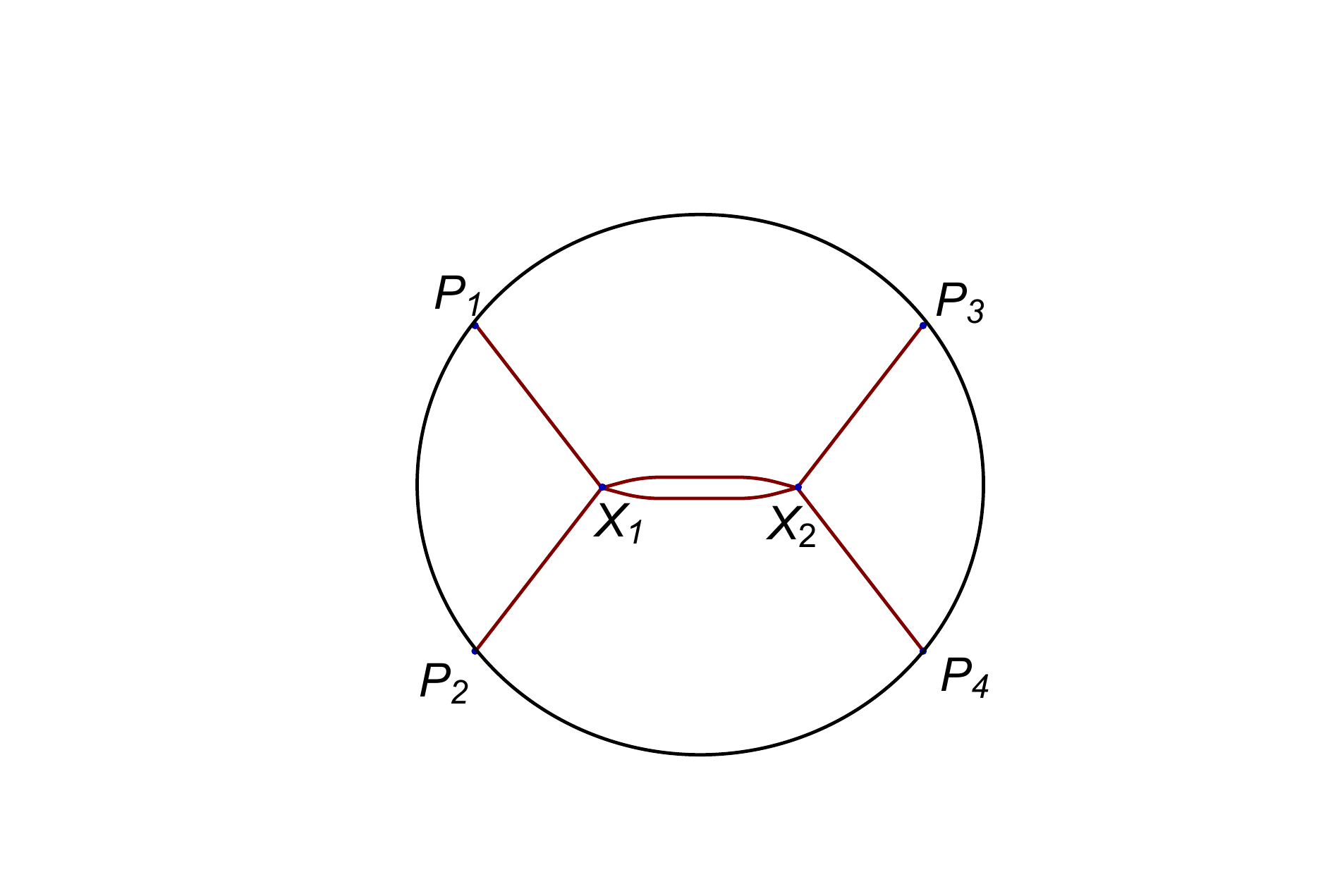}
  \caption{Bubble Witten diagram.}\label{bubble1}
\end{figure}
The bubble Witten diagram corresponds to a similar expression as \eqref{exchange} but with two bulk-to-bulk propagators connecting the same points $X_1,\,X_2$ instead of one, as shown in figure \ref{bubble1}. In practice, we can think of the product of the two bulk-to-bulk propagators that depend on the same $u$, as a single bulk-to-bulk propagator  of the form \eqref{normalbulk} but with two $c-$integrations  and two  $\gamma-$integrations. That particularity allow us to do a single expansion of the type \eqref{mellinbarnes} on the product, instead of one for each propagator.
The amplitude associated to this diagram is given by
\be\label{bubble}
\begin{split}
&{\cal A}^{\mathrlap{\vee}\wedge}_4 (P_1, P_2, P_3, P_4) = \\
&g^2 \int_{AdS}\dd X_1 \dd X_2G^{\Delta_1}_{\d B}(P_1, X_1)  G^{\Delta_2}_{\d B}(P_2, X_2)  G^{\Delta}_{B B}(X_1, X_2) G^{\Delta'}_{B B}(X_1, X_2) G^{\Delta_3}_{B\d}(X_2, P_3) G^{\Delta_4}_{B\d}(X_2, P_4)\; .
\end{split}
\ee
 Replacing the definitions for the bulk-to-boundary propagators, the representation \eqref{normalbulk} and using \eqref{mellinbarnes} we get,
  \be\label{bubble2}
\begin{split}
&{\cal A}^{\mathrlap{\vee}\wedge}_4 (P_i) =g^2 \prod_{i=1}^4{\cal C}_{\Delta_i} \prod_{j=1}^2\int_{-i\infty}^{i\infty}{\dd c_j\,\dd \gamma_j \dd f\over (2\pi i)^4}\\
&{\Gamma(h+c_j-\gamma_j)\Gamma(h-c_j-\gamma_j)\Gamma(\gamma_j)\Gamma(h-\gamma_j)\over \Gamma(c_j)\Gamma(-c_j)((\Delta-h)^2-c_1^2)((\Delta'-h)^2-c_2^2) \Gamma(2h-2\gamma_j)}
{\Gamma(\gamma_1+\gamma_2+f)\Gamma(-f)(-2)^f\over \Gamma(\gamma_1+\gamma_2)}\\
&
 \int_{AdS}{\dd X_1 \dd X_2\over (-2P_1\cdot X_1)^{\Delta_1}  (-2P_2\cdot X_1)^{\Delta_2} (-2(X_1\cdot X_2))^{\gamma_j+f_j}  (-2P_3\cdot X_2)^{\Delta_3} (-2P_4\cdot X_2)^{\Delta_4}}\,.
\end{split}
\ee
we can notice that the computation should follows the same lines as the exchange diagram in section above with $\Delta_5$ in \eqref{exchange3} replaced by $\Delta_5=\gamma_1+\gamma_2+f$, leading us to the following representation of the corresponding Mellin amplitude,
\be
\begin{split}
&M^{\mathrlap{\vee}\wedge}_4=g^2\prod_{j=1}^2\int_{-i\infty}^{i\infty}{\dd c_j\,\dd \gamma_j\over (2\pi i)^4}\\
&{\Gamma(h+c_j-\gamma_j)\Gamma(h-c_j-\gamma_j)\Gamma(\gamma_j)\Gamma(h-\gamma_j)\over \Gamma(c_j)\Gamma(-c_j)((\Delta-h)^2-c_1^2)((\Delta'-h)^2-c_2^2) \Gamma(2h-2\gamma_j)}
{1\over \Gamma(\gamma_1+\gamma_2)}\\
&~~
\frac{\Gamma \left(\frac{\gamma_1+\gamma_2 -s}{2}\right) \Gamma \left(\frac{-2 h+\gamma_1+\gamma_2+\Delta _1+\Delta _2}{2}\right) \Gamma \left(\frac{-2 h+\gamma_1+\gamma_2
   +\Delta _3+\Delta _4}{2} \right) }{\Gamma \left(\frac{-2 h+\Delta _1+\Delta _2+\Delta _3+\Delta _4}{2} \right) \Gamma \left(\frac{-s+\Delta
   _1+\Delta _2}{2} \right) \Gamma \left(\frac{-s+\Delta _3+\Delta _4}{2}\right)}\,.
\end{split}
\ee
\section{One-loop vertex correction}
\begin{figure}[h]
	\centering
  \includegraphics[width=6.0cm]{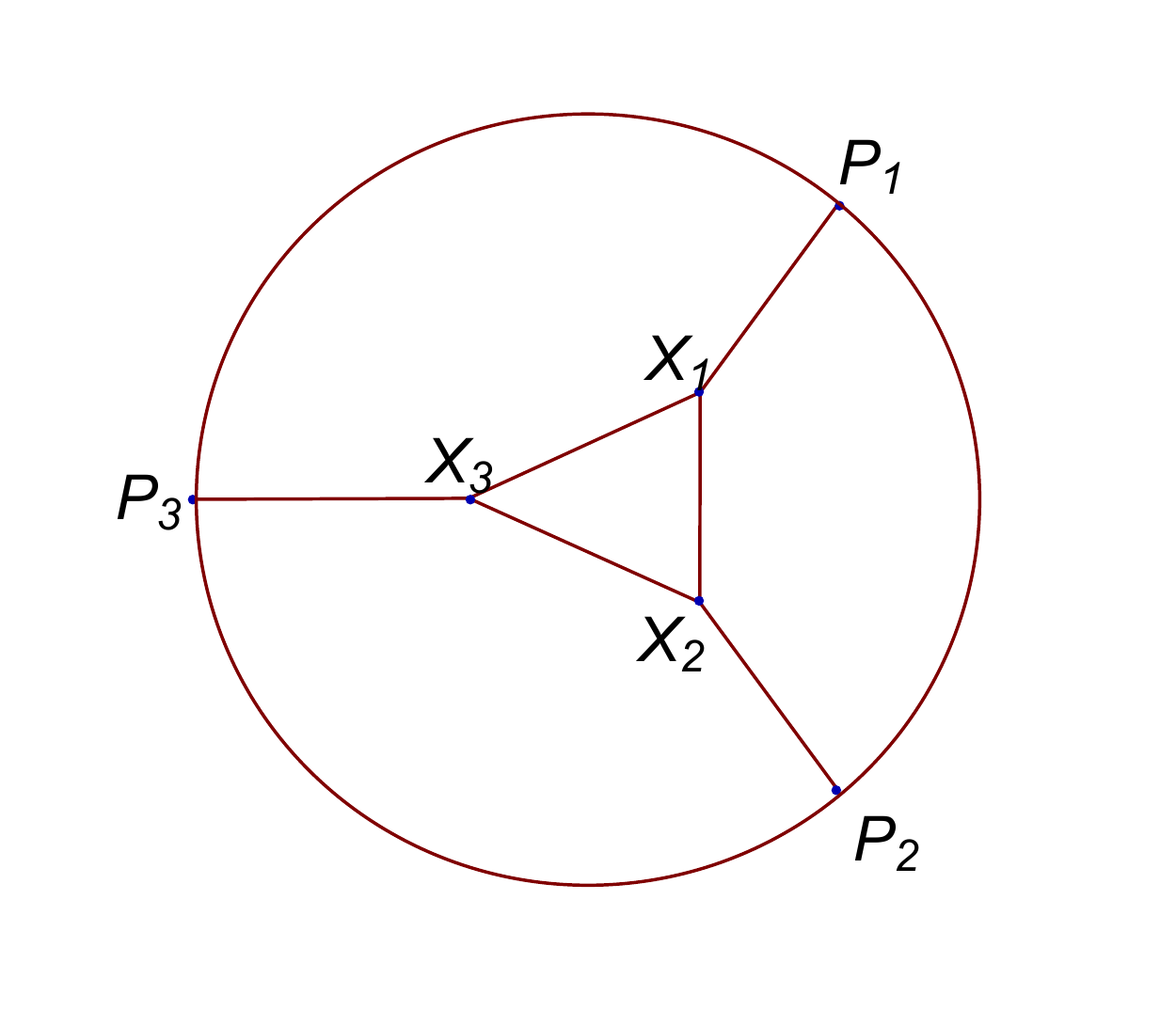}\caption{ One loop Vertex correction.}\label{vertex}
\end{figure} 
Before moving onto the remaining loop corrections to the four point scattering amplitude, let us consider the one-loop correction to the three point function,  represented in figure \ref{vertex} and given by,
\be\label{vertex1}
{\cal A}^{\rhd}_3 (P_i) = 
g^3\prod_{i=1}^{3\,{\rm Mod}3}\int_{AdS}\dd X_i\,G^{\Delta_i}_{\d B}(P_i, X_i)  G^{\Delta_{ii+1}}_{B B}(X_i, X_{i+1}) \,.
\ee
Introducing Schwinger parameters in the usual way we have,
\be
\begin{split}
&{\cal A}^{\rhd}_3(P_i) = 
g^3\prod_{i=1}^{3\,{\rm Mod}3}\,\widehat{G}^{\Delta_{ii+1}}_{B B}(c_{i,i+1},\,\gamma_{i,i+1},\,f_{i,i+1})\\
&~~~~\times\prod_{j=1}^7\int_0^{\infty}{\dd t_j\over t_j\,\Gamma(\Delta_j)} t_j^{\Delta_j}\int_{AdS}\dd X_i\e^{2(t_1P_1+t_5X_2+t_7X_3)\cdot X_1}\e^{2(t_2P_2+t_6X_3)\cdot X_2}\e^{2(t_3P_3)\cdot X_3}\,,
\end{split}
\ee
where we have redefined the conformal weights  of the bulk-to-bulk propagators in \eqref{vertex1} as  $\Delta_5\equiv\Delta_{12}=\gamma_{12}+f_{12},\,~ \Delta_6\equiv\Delta_{23}=\gamma_{23}+f_{23},\,~\Delta_7\equiv\Delta_{13}=\gamma_{31}+f_{31}$.
We then proceed to integrate the bulk coordinates in the exact same manner as in the previous cases to get 
\be\label{mellinvertex}
\begin{split}
&{\cal M}^{\rhd}_3(P_i) = 
 \prod_{i=1}^{3\,{\rm Mod}3}\int_{-i\infty}^{i\infty}{\dd c_{i\,i+1}\,\dd \gamma_{i\,i+1} \dd  f_{i\,i+1}\over (2\pi i)^4}{\Gamma(h+c_{i\,i+1}-\gamma_{i\,i+1})\Gamma(h-c_{i\,i+1}-\gamma_{i\,i+1})\over \Gamma(c_{i\,i+1})\Gamma(-c_{i\,i+1})((\Delta-h)^2-c_{i\,i+1}^2)}\\
 &{\Gamma(h-\gamma_{i\,i+1})\Gamma(-f_{i\,i+1})(-2)^f_{i\,i+1}\over \Gamma(2h-2\gamma_{i\,i+1})}\\
 &\Gamma\left({\Delta_1+\gamma_{12}+f_{12}+\gamma_{31}+f_{31}-2h\over 2}\right)\Gamma\left({\Delta_2+\gamma_{12}+f_{12}+\gamma_{23}+f_{23}-2h\over 2}\right)\\
 &\Gamma\left({\Delta_3+\gamma_{23}+f_{23}+\gamma_{31}+f_{31}-2h\over 2}\right)\\
 &\prod_{j=5}^7\int{\dd t_j\over t_j}\, t_j^{\Delta_j}\,{T_7^{-\delta_{12}}\,T_6^{-\delta_{13}}\,T_5^{-\delta_{23}}\,(1+t_7^2+T_5^2)^{-{\Delta_3+\gamma_{23}+f_{23}+\gamma_{31}+f_{31}-2h\over 2}}\over(1+t_5^2)^{{\Delta_2+\gamma_{12}+f_{12}+\gamma_{23}+f_{23}-2h\over 2}}}
\end{split}
\ee
where for convenience of notation we have defined
$T_5=t_5t_7+t_6,\, T_6=t_5T_5+t_7,\,T_7=T_6T_5$. Notice that in this case, the Mellin amplitude does not depend on any kinematical degrees of freedom, or in other words it is just a constant, since the solution of the system of equation \eqref{kinematics} is simply,
\be
\delta_{12}={\Delta_1+\Delta_2-\Delta_3\over 2},\,~\delta_{13}={\Delta_1+\Delta_3-\Delta_2\over 2},\,~\delta_{23}={\Delta_2+\Delta_3-\Delta_1\over 2}\,.
\ee 
This is just the reflection of the fact that the three-point function is fixed by conformal invariance up to a constant, so what we are really computing here is the loop correction of the three point structure constant.

Let us focus now on the integral on the last line of \eqref{mellinvertex},
\be\label{tvertex}
\begin{split}
&\int{\dd t_5\over t_5}{\dd T_5}{\dd t_7\over t_7}\,t_5^{\Delta_5}(T_5-t_5t_7)^{\Delta_6-1}\,t_7^{\Delta_7}\,\\
&\times{(t_5T_5+t_7)^{-\delta_{12}-\delta_{13}}\,T_5^{-\delta_{12}-\delta_{23}}\,(1+t_7^2+T_5^2)^{-{\Delta_3+\gamma_{23}+f_{23}+\gamma_{31}+f_{31}-2h\over 2}}\over(1+t_5^2)^{{\Delta_2+\gamma_{12}+f_{12}+\gamma_{23}+f_{23}-2h\over 2}}}
\end{split}
\ee
where we have replaced $t_6$ by $T_5$. This looks like a difficult integral to perform, let alone in a clean compact way, so we choose at this point to perform a brute force approach. In order to do so, we expand the factor $(1+t_7^2+T_5^2)$ in the numerator \eqref{tvertex} by means of applying the Mellin-Barnes transform \eqref{mellinbarnes} one more time,
\be\label{tvertex2}
\begin{split}
&\int_{-i\infty}^{i\infty}{\dd\ell\over 2\pi i}{\Gamma(a+\ell)\Gamma(-\ell)\over \Gamma(a)}I^{\rhd}_S\,,
\end{split}
\ee
where we have defined the new integral,
\be
I^{\rhd}_S\equiv\int{\dd t_5\over t_5}{\dd T_5}{\dd t_7\over t_7}\,t_5^{\Delta_5}(T_5-t_5t_7)^{\Delta_6-1}\,t_7^{\Delta_7} {(t_5T_5+t_7)^{-\delta_{12}-\delta_{13}}\,T_5^{2\ell-\delta_{12}-\delta_{23}}\over(1+t_5^2)^{b}(1+t_7^2)^{a+\ell}}
\ee
and $a={{\Delta_3+\gamma_{23}+f_{23}+\gamma_{31}+f_{31}-2h\over 2}}$, $b={\Delta_2+\gamma_{12}+f_{12}+\gamma_{23}+f_{23}-2h\over 2}$. With the help of Mathematica \cite{Mathematica} this expansion allow us to represent the integral over all the remaining Schwinger parameters in terms of a lengthy combination of  Hypergeometric functions, as is shown in Appendix equation \eqref{onepagetriangle}. However, we will discuss an approach to deal with a more general family of integrals in section 7 which includes the one above.
 
\section{Triangle Witten diagram }
\begin{figure}[h]
	\centering
  \includegraphics[width=6.0cm]{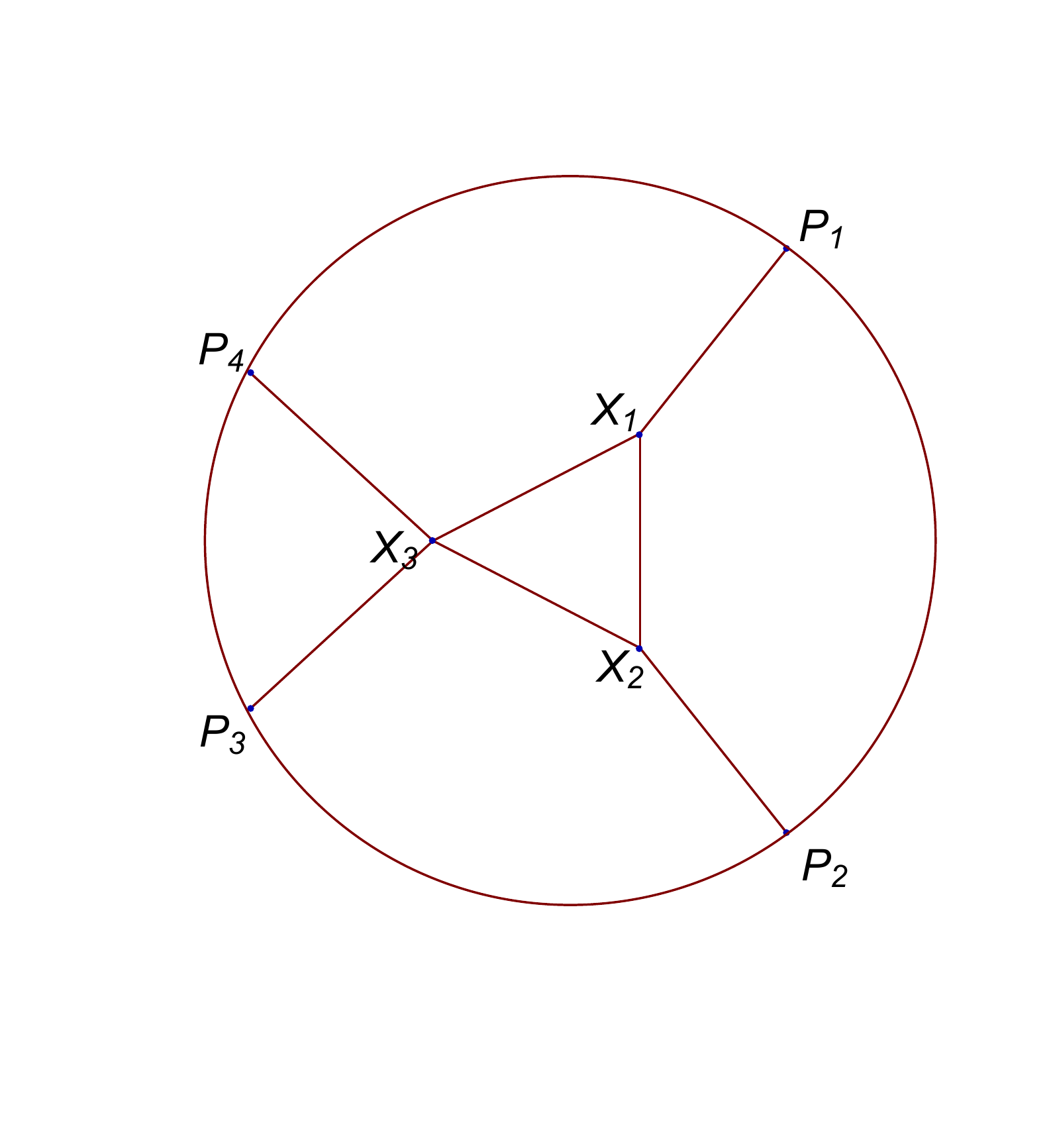}\caption{ One loopTriangle.}\label{fig2}
\end{figure} 
The 1-loop four-particle triangle Witten diagram is represented by the figure \ref{fig2}  and can be written in the form,
\be
{\cal A}^{\rhd} (P_i) = 
g^3\prod_{i=1}^3\int_{AdS}\dd X_iG^{\Delta_i}_{\d B}(P_i, X_i)  G^{\Delta_4}_{B\d}(P_4,X_3)\prod_{i<j}^3G^{\Delta_{ij}}_{B B}(X_i, X_j)  \,.
\ee
We use again the representation \eqref{mybulk} and start by introducing Schwinger parameters to exponenciate the propagators,
\be
\begin{split}
{\cal A}^{\rhd} (P_i)& = 
g^3\prod_{i=1}^4{\cal C}_{\Delta_i} \prod_{i<j}^3\int_{-i\infty}^{i\infty}{\dd c\,\dd \gamma_{ij} \dd f_{ij}\over (2\pi i)^3}\widehat{G}^{\Delta_{ij}}_{BB}(c,\gamma_{ij}, f_{ij})\\
&\int^{\infty}_{0}\prod_{i=1}^7{\dd t_i\over t_i}{t_i^{\Delta_i}\over \Gamma(\Delta_i)}\prod_{i=1}^3\int_{AdS}\dd X_i\e^{2(t_1P_1+t_5X_2+t_7 X_3)\cdot X_1}\e^{2(t_2P_2+t_6X_3)\cdot X_2} \e^{2(t_3P_3+t_4P_4)\cdot X_3} \,.
\end{split}
\ee
where for convenience of notation in the last line we have redefined the conformal weights of the bulk-to-bulk propagators as,
\be
\Delta_5\equiv\Delta_{12}=\gamma_{12}+f_{12},\,~~\Delta_6\equiv\Delta_{23}=\gamma_{23}+f_{23},\,~~\Delta_7\equiv\Delta_{13}=\gamma_{13}+f_{13}\,. 
\ee
By taking $Q_1=(t_1P_1+t_5X_2+t_7 X_3)$, using \eqref{bulkintegral}  and scaling $(t_1,t_5,t_7)\to \sqrt{z_1}(t_1,t_5,t_7)$ we can write,
\be
\begin{split}
{\cal A}^{\rhd} (P_i)& = 
g^3\prod_{i=1}^4{\cal C}_{\Delta_i} \prod_{i<j}^3\int_{-i\infty}^{i\infty}{\dd c_{ij}\,\dd \gamma_{ij} \dd f_{ij}\over (2\pi i)^3}\widehat{G}^{\Delta_{ij}}_{BB}(c_{ij},\gamma_{ij}, f_{ij})\\
&\int^{\infty}_{0}\prod_{i=1}^7{\dd t_i\over t_i}{t_i^{\Delta_i}\over \Gamma(\Delta_i)}\e^{Q_1^2}\prod_{i=2}^3\int_{AdS}\dd X_i\e^{2(t_2P_2+t_6X_3)\cdot X_2} \e^{2(t_3P_3+t_4P_4)\cdot X_3}\\
&~~~\times\Gamma\left({\Delta_1+\gamma_{12}+f_{12}+\gamma_{13}+f_{13}-2h\over 2}\right) \,.
\end{split}
\ee
Following the same steps for the integration over $X_2$ and $X_3$ we end up with the expression,
\be\label{triangle2}
\begin{split}
&{\cal A}^{\rhd} (P_i) = 
g^3\prod_{i=1}^4{\cal C}_{\Delta_i} \prod_{i<j}^3\int_{-i\infty}^{i\infty}{\dd c\,\dd \gamma_{ij} \dd f_{ij}\over (2\pi i)^3}\widehat{G}^{\Delta_{ij}}_{BB}(c_{ij},\gamma_{ij}, f_{ij})\,\e^{2t_1t_2 P_1\cdot P_2}\\
&\int^{\infty}_{0}\prod_{i=1}^7{\dd t_i\over t_i}{t_i^{\Delta_i}\over \Gamma(\Delta_i)}\,\e^{Q_3^2}\,\Gamma\left({\Delta_1+\gamma_{12}+f_{12}+\gamma_{13}+f_{13}-2h\over 2}\right)\Gamma\left({\Delta_2+\gamma_{12}+f_{12}+\gamma_{23}+f_{23}-2h\over 2}\right) \\
&\int_0^{\infty}{\dd z_3\over z_3}z_3^{\Delta_3+\Delta_4+\gamma_{23}+f_{23}+\gamma_{13}+f_{13}-2h\over 2}{\e^{-z_3(1+(t_5t_7+t_6)^2)}\over (t_5^2+z_3t_7^2)^{\Delta_2+\gamma_{12}+f_{12}+\gamma_{23}+f_{23}-2h\over 2}}\,,
\end{split}
\ee
with $Q_3=(t_7+t_5(t_5t_7+t_6))t_1 P_1+(t_5t_7+t_6)t_2P_2+t_3 P_3+t_4P_4$.

As we have pointed out already, we can notice from this expression that the increasing complexity of the integral over $z_i$ comes from the $t_i^2$ factors left behind by the on-shell condition $X_i^2=-1$. The integral over $z_3$ can be represented as a combination of Hypergeometric functions, but since is somewhat lengthy we
choose instead to introduce yet another Mellin parameter to trivialize this integral by Mellin transform the exponent at the last line of \eqref{triangle2},
\be
\begin{split}
&\int_{-i\infty}^{i\infty}{\dd g\over (2\pi i)}\Gamma(-g)\int_0^{\infty}{\dd z_3\over z_3}z_3^{\Delta_3+\Delta_4+\gamma_{23}+f_{23}+\gamma_{13}+f_{13}+g-2h\over 2}{(1+(t_5t_7+t_6)^2)^g\over (t_5^2+z_3t_7^2)^{\Delta_2+\gamma_{12}+f_{12}+\gamma_{23}+f_{23}-2h\over 2}}\\
&=\int_{-i\infty}^{i\infty}{\dd g\over (2\pi i)}\Gamma(-g)\frac{\Gamma \left(\frac{-g+f_{12}-f_{13}+\gamma _{12}-\gamma _{13}+\Delta _2-\Delta _3-\Delta _4}{2} \right)\Gamma \left(\frac{g-2 h+f_{13}+f_{23}+\gamma _{13}+\gamma _{23}+\Delta _3+\Delta
   _4}{2} \right)}{\Gamma
   \left(\frac{-2 h+f_{12}+f_{23}+\gamma _{12}+\gamma _{23}+\Delta _2}{2}\right)}  \\
   &~~~ t_5^{g-\gamma _{12}-f_{12}-\gamma _{13}-f_{13}-\Delta _2+\Delta_3-\Delta_4}t_7^{-\gamma _{13}-\gamma _{23}-\Delta _3-\Delta _4-f_{13}-f_{23}-g+2 h}(1+(t_5t_7+t_6)^2)^g\,.
\end{split}
\ee
Putting it back in \eqref{triangle2} and using Symanzik identity \eqref{Syma} we have that the Mellin amplitude associated to the one-loop triangle can be written as,
\be
\begin{split}\label{trianglemellin}
&M^{\rhd} (s,\,t) = 
g^3\prod_{i=1}^4{\cal C}_{\Delta_i} \prod_{i<j}^3\int_{-i\infty}^{i\infty}{\dd c\,\dd \gamma_{ij} \dd f_{ij}\dd g\over (2\pi i)^4}\Gamma(-g)\,\widehat{G}^{\Delta_{ij}}_{BB}(c_{ij},\gamma_{ij}, f_{ij})\\
&\Gamma\left({\Delta_1+\gamma_{12}+f_{12}+\gamma_{13}+f_{13}-2h\over 2}\right)\Gamma\left({\Delta_2+\gamma_{12}+f_{12}+\gamma_{23}+f_{23}-2h\over 2}\right) \\
&\frac{\Gamma \left(\frac{-g+f_{12}-f_{13}+\gamma _{12}-\gamma _{13}+\Delta _2-\Delta _3-\Delta _4}{2} \right)\Gamma \left(\frac{g-2 h+f_{13}+f_{23}+\gamma _{13}+\gamma _{23}+\Delta _3+\Delta
   _4}{2} \right)}{\Gamma
   \left(\frac{-2 h+f_{12}+f_{23}+\gamma _{12}+\gamma _{23}+\Delta _2}{2}\right)}  \\
   &\int^{\infty}_{0}\prod_{i=5}^7{\dd t_i\over t_i}{1\over \Gamma(\Delta_i)}\,t_5^{g+\gamma _{13}+f_{13}-\Delta _2+\Delta_3+\Delta_4}\,t_6^{\gamma _{23}+f_{23}}\,t_7^{-\gamma _{23}-f_{23}-\Delta _3-\Delta _4-g+2 h}(1+T_5^2)^g\\
   &~~~~~~T_7^{-\delta_{12}}T_6^{-\delta_{13}-\delta_{14}}T_5^{-\delta_{23}-\delta_{24}}\,,
\end{split}
\ee
where we have defined for convenience on the notation,  $T_5=(t_5 t_7+t_6),\,T_6=(t_7+t_5T_5),\,T_7=(t_5+T_6T_5)$.  Interestingly, the integration over the remaining Schwinger parameters does not depend on $\gamma _{12}$ and $f_{12}$. 
Below we will propose a way to treat the remaining integration over Schwinger parameters on the last line of \eqref{trianglemellin}, but before doing so, let us discuss first the box Witten diagram in the next section.
\section{Box Witten diagram}
\begin{figure}[h]
	\centering
  \includegraphics[width=6.0cm]{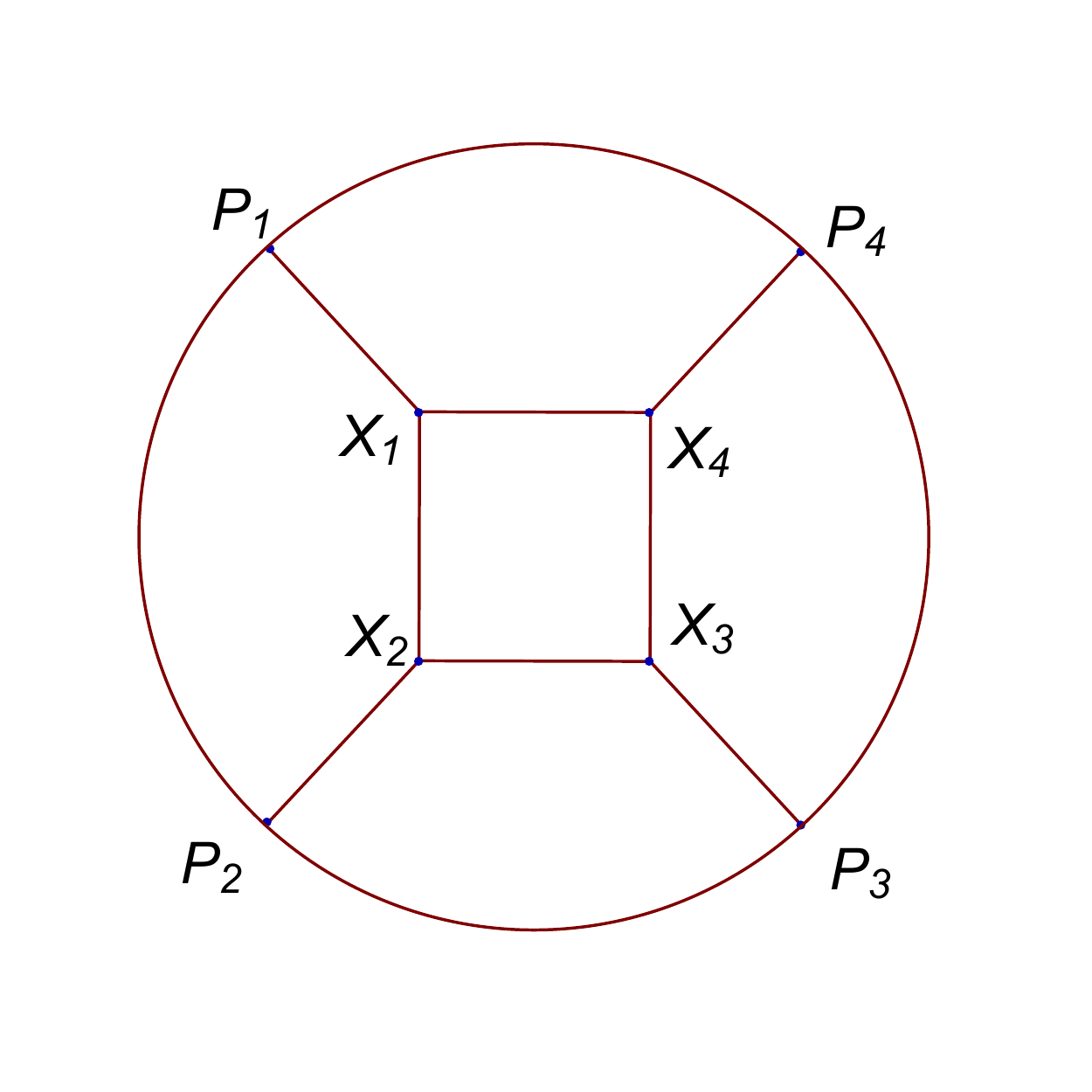}\caption{ One loop box.}\label{box}
\end{figure} 
The 1-loop four-particle box Witten diagram is represented by the figure \ref{box}  and can be written in the form, 
\be
{\cal A}^{\Box} (P_i) = 
g^3\prod_{i=1}^{4\,{\rm Mod}4}\int_{AdS}\dd X_i\,G^{\Delta_i}_{\d B}(P_i, X_i)  G^{\Delta_{ii+1}}_{B B}(X_i, X_{i+1}) \,.
\ee
Introducing Schwinger parameters in the usual way we have,
\be
\begin{split}
&{\cal A}^{\Box} (P_i) = 
g^4\prod_{i=1}^{4\,{\rm Mod}4}\,\widehat{G}^{\Delta_{ii+1}}_{B B}(c_{i,i+1},\,\gamma_{i,i+1},\,f_{i,i+1})\\
&~~~~\times\prod_{j=1}^8{\dd t_j\over t_j\,\Gamma(\Delta_j)} t_j^{\Delta_j}\int_{AdS}\dd X_i\e^{2(t_1P_1+t_5X_2+t_8X_4)\cdot X_1}\e^{2(t_2P_2+t_6X_3)\cdot X_2}\e^{2(t_3P_3+t_7X_4)\cdot X_3}\e^{2t_4P_4\cdot X_4}\,.
\end{split}
\ee
where we have defined $\Delta_5=\gamma_{12}+f_{12},\, \Delta_6=\gamma_{23}+f_{23},\,\Delta_7=\gamma_{34}+f_{34}$ and $\Delta_8=\gamma_{41}+f_{41}$. We then proceed to integrate the bulk coordinates in the exact same manner as in the previous cases to get 
\be\label{mellinbox}
\begin{split}
&M^{\Box}(s,t)=
 \prod_{i=1}^{4\,{\rm Mod}4}\int_{-i\infty}^{i\infty}{\dd c_{i\,i+1}\,\dd \gamma_{i\,i+1} \dd  f_{i\,i+1}\over (2\pi i)^4}{\Gamma(h+c_{i\,i+1}-\gamma_{i\,i+1})\Gamma(h-c_{i\,i+1}-\gamma_{i\,i+1})\over \Gamma(c_{i\,i+1})\Gamma(-c_{i\,i+1})((\Delta-h)^2-c_{i\,i+1}^2)}\\
 &{\Gamma(h-\gamma_{i\,i+1})\Gamma(-f_{i\,i+1})(-2)^f\over \Gamma(2h-2\gamma_{i\,i+1})}\\
 &\Gamma\left({\Delta_1+\gamma_{12}+f_{12}+\gamma_{41}+f_{41}-2h\over 2}\right)\Gamma\left({\Delta_2+\gamma_{12}+f_{12}+\gamma_{23}+f_{23}-2h\over 2}\right)\\
 &\Gamma\left({\Delta_3+\gamma_{23}+f_{23}+\gamma_{34}+f_{34}-2h\over 2}\right)\Gamma\left({\Delta_4+\gamma_{34}+f_{34}+\gamma_{41}+f_{41}-2h\over 2}\right)\\
 &\prod_{j=5}^8{\dd t_j\over t_j}\, t_j^{\Delta_j}\,{(1+t_8^2(1+t_5^2)+T_5^2)^{-{\Delta_4+\gamma_{34}+f_{34}+\gamma_{41}+f_{41}-2h\over 2}}\over(1+t_5^2)^{{\Delta_2+\gamma_{12}+f_{12}+\gamma_{23}+f_{23}-2h\over 2}}(1+t_6^2)^{{\Delta_3+\gamma_{23}+f_{23}+\gamma_{34}+f_{34}-2h\over 2}}}\\
 &~~~~~\times T_{10}^{-\delta_{12}}T_9^{-\delta_{13}}T_8^{-\delta_{23}}\,T_7^{-\delta_{14}}\,T_6^{-\delta_{24}}\,T_5^{-\delta_{34}}
\end{split}
\ee
where in order to simplify the notation we have defined,

\be
\begin{split}
T_5&=t_5t_8t_6+t_7,\,~ T_6=t_6T_5+t_5t_8,\,~T_7=t_5T_6+t_8,\,~T_8=t_6+T_5T_6,\\
T_9&=t_5t_6+T_7T_5,\,~T_{10}=t_5(1+t_6^2)+T_7T_6
\end{split}
\ee 

At this point some comments are in order. In the last two examples we have been able to write the one loop Mellin representation of the four-point triangle and box Witten diagrams in terms of multiple contour integrals and  some remaining integrals over Schwinger parameters. Let us assume for a moment that we are able to evaluate all the integrations over the Schwinger parameters, then we still will be left with a multiple contour integration of Mellin-barnes type. Even thought performing this multiple integrals can still be tricky, is not hard to convince ourselves that those contour integrations are in principle much more simple than the type of integrals involving Schwinger parameters, since they are picking up residues only over (a rational function of) gamma functions, which only have simple poles. Therefore, the final result from such a multiple contour  integration is expected to be, in general, an infinite sum of all those residues, without any derivatives. In fact, Hypergeometric functions are represented as contour integrals  over rational functions of gamma functions, similar to the ones we have on the given examples. \footnote{Actually, the contour integrals we have here are more like Hypergeometric functions evaluated at $z=1$, namely, $_pF_q(\{a\},\{b\}| 1)$, which are simpler and have simplifying representations.}
There is still the possibility of having singularities coming from poles trapping and  pinching the integration contour, but on physical grounds those are the singularities expected to produce the poles in the Mellin amplitude.

In order to the above argument to hold, we should hope to be able to write the leftover integration over Schwinger parameters as a rational function of gamma functions or even as a similar contour integral over rational functions of gamma functions. Even though we did not manage to explicitly do it in this work, in the next section we present a proposal to treat the aforementioned integrals as contour integrals. 

Before proceed to the next section, it is worth to notice that we also can deal with the remaining Schwinger parameters by iteratively apply formula \eqref{mellinbarnes} to convert those integrals at the last line at \eqref{trianglemellin} and \eqref{mellinbox} as contour integrations over Beta functions. However, on the one hand we would like to have a representation with the minimal amount of integrations, and this procedure requires the introduction of at least an additional contour integral for each polynomial $T_i$, since essentially formula \eqref{mellinbarnes} would be used to ``binomial' expand each of them. On the other hand, this procedure seems quite arbitrary since there is many different ways to choose the factors playing the role of $A$ and $B$ in \eqref{mellinbarnes} and we don't have a criteria to pick the more convenient choice.
\footnote{We thank to  Ellis Ye Yuan for discussions on this point.}
\section{Proposal on how to treat the integration over the remaining Schwinger parameters.}
In this section we would like to propose a procedure to deal with the integration of Schwinger parameters associated to the bulk-to-bulk propagators that were left behind after using the Symanzik formula \eqref{Syma}. The type of integrals we would like to consider have the schematic form,
\be\label{remaining}
{\cal I}_{n_{int}}\equiv\int^{\infty}_{0}\prod_{i>n_{\rm ext}}^{n_{\rm ext}+n_{\rm int}}{\dd t_i}\,R(\{t_i\})\prod_{i<j}^{n_{ext}}\left({\over}T_{ij}(\{t_i\})\right)^{-\delta_{ij}}\,,
\ee
where $n_{\rm ext},\,n_{\rm int}$ are the number of external particles (attached to the boundary of AdS)\footnote{In this paper we are only focused on $n_{\rm ext}=4$, but we prefer to keep calling it $n_{\rm ext}$ since we think most of the discussion in this section applies equally to the more general $n-$gon case.} and the number of internal particles (or bulk-to-bulk propagators) respectively, $\{t_i\}$ denotes the set of integration variables, $T_i(\{t_i\})$ are polynomials with integer coefficients and $R(\{t_i\})$ is in general rational function. 

Let us start illustrating the procedure with an over simplified example, namely a  beta-like function with a negative parameter,
\be
{\cal I}_0=
\int^{\infty}_{0}{\dd t\over t}\,t^{\alpha}\,(1+t)^{\beta}\,.
\ee
We take the exponents $\alpha$  as negative integers and at the end we  analytic continue them to complex values.
We first Wick rotate  the integration contour to the imaginary axes and pick up the residues sitting at the positive positive quarter of the complex plane. Lets us start performing the integration by using the Cauchy integration formula  picking up the poles at $t=0$,
\be
{\cal I}_0={\Gamma(\beta+1)\over\Gamma(-\alpha+1) \Gamma(\beta+\alpha+1)}\,.
\ee
As desired, the integration over this hypothetical Schwinger parameter is given by a rational function of gamma functions. Notice as well that in this form $\alpha$ can be analytically continue to generic complex values. For example, it can be related to the more familiar Beta function by analitically continue it to $\mathbb{R}(\alpha)>0$  \& $\alpha\notin \mathbb{Z}$, by means of the identity,
\be
 \Gamma(z)\Gamma(1-z)={\pi\over {\rm sin}(\pi z)},\quad z\notin \mathbb{Z}\,,
\ee
leading us to 
\be
{\cal I}_0={\rm B}(\alpha, \beta+1){{\rm sin}\pi (-\alpha+1)\over \pi}\,.
\ee
Maybe a more familiar example is given by writing the  Beta function as a contour integral over the Pochhammer contour,
\be
(1-\e^{2\pi \alpha}) (1-\e^{2\pi \beta})B(\alpha,\beta) =\int_{\cal C}t^{\alpha-1}(1-t)^{\beta-1}\,,
\ee
which corresponds to the analytic continuation of the Beta function for all values of $\alpha$ and $\beta$.

\subsection{Sketching the general approach}
In this section we would like to sketch, without getting into details, a generalization to the procedure illustrated above  to deal with the more general integral \eqref{remaining}. As in the previous example we consider the parameters $\delta_{ij} $ as positive integers and we start by Wick rotate the integration contour to pick up the residues at the roots of the polynomials in the denominator of \eqref{remaining}, including the ones in the function $R(t_i)$, analytically continuing the result to more general values at the end. 


Before moving on, is interesting to notice that we can homogenize every polynomial by the introduction of a delta function, namely,
\be 
1=\int dt \delta(t-1)\,
\ee
such as that on every monomial composing the polynomial we include as many factors of $t$ as necessary to make the whole polynomial homogeneous. For example, the integral over the Schwinger parameters at \eqref{trianglemellin} can be rewritten as,
\be\begin{split}
  &\int dt \int^{\infty}_{0}\prod_{i=5}^7{\dd t_i\over t_i}{\delta(t-1)\over \Gamma(\Delta_i)}\,t_5^{g+\gamma _{13}+f_{13}-\Delta _2+\Delta_3+\Delta_4}\,t_6^{\gamma _{23}+f_{23}}\,t_7^{-\gamma _{23}-f_{23}-\Delta _3-\Delta _4-g+2 h}(1+\hat{T}_5^2)^g\\
   &~~~~~~\hat{T}_7^{-\delta_{12}}\hat{T}_6^{-\delta_{13}-\delta_{14}}\hat{T}_5^{-\delta_{23}-\delta_{24}}\,,
\end{split}
\ee
where we have redefined the homogenized polynomials
\be \hat{T}_5=(t_5 t_7+t_6\, t),\,~~\hat{T}_6=(t_7\, t^2+t_5T_5),\,~~\hat{T}_7=(t_5\, t^4+T_6T_5)\,.\ee  
Even more, the polynomials $(\hat{T}_i)^{\rho_i}$ are still homogeneous with the same set of zeros as $\hat{T} _i$ but with higher multiplicity. Therefore the problem of finding the zero locus of the set of polynomials can be rephrased in terms of the problem of looking for the intersections of algebraic curves in ${\mathbb P}^n$. This can be an interesting point of view, linking the problem at hand with similar approaches to the flat space S-matrix, where the scattering amplitudes elements are written as contour integrals on the support of the intersection of rational curves \cite{ArkaniHamed:2009dg}, as in the present case. It would be interesting to  explore that point of view in  future studies.

Let us return to our goal in this section by rewrite \eqref{remaining} as,
\be\label{remaining2}
{\cal I}_{n_{int}}\equiv\int_{\cal C}\prod_{i=1}^{\cal I}{\dd t_i}\,{G(\{t_i\})\over \prod_{j=1}^{m}P_{j}(\{t_i\})}\,,
\ee
where we have explicitly displayed all the polynomial dependence on the denominator, such as that the number $m$ of polynomials in the denominator  is  larger than the number of index couples $\{ij\}$, i.e, $m>(n_{\rm ext}(n_{\rm ext}-1)/2)\,~$, and $G(\{t_i\})$ is assume to be an holomorphic function inside the contour ${\cal C}$, given by the cycle $C_{\cal C}(\{t_i\})=\{\{t_i\}:P_j(\{t_i\})=\epsilon\}$ with $\epsilon$  small. ${\cal I}$  is the number of integration variables. Notice that along the way we have defined some of the polynomials to be,
\be 
P_{j}(\{t_i\})=\left({\over}T_{j}(\{t_i\})\right)^{\delta_{j}},\,~~~~~\text{for}~~~~~ j=1,\cdots,{n_{\rm ext}(n_{\rm ext}-1)\over2}\,.
\ee
 In this form, we can define local residues associated to the common roots $\tilde{t}_i$ of a subset of polynomials $\{P_{q_1},\cdots P_{q_{\cal I}}\}\in\{P_{1},\cdots P_{m}\}$ in the following way \cite{Griffiths, 1994alg.geom:4011C}. Take,
\be
H(\{\tilde{t}_i\})_{q_1,\cdots,q_{\cal I}}= {G(\{\tilde{t}_i\})\over \prod_{j\neq {q_1,\cdots,q_{\cal I}}}P_{j}(\{\tilde{t}_i\})}
\ee
then if the  Jacobian $J(\{\tilde{t}_i\})$,
\be J(\{\tilde{t}_i\})={\rm Det}\left({\partial (P_{q_1},\cdots P_{q_{\cal I}})\over \partial(\{t_i\})}\right)(\{\tilde{t}_i\})\ee
is not degenerate, i.e, if $ J(\{\tilde{t}_i\})\neq 0$, then a local residue can be defined as,
\be\label{nondegresidue}
{\rm Res}_{\{P_{q_1},\cdots P_{q_{\cal I}}\}}\left({G(\{t_i\})\over \prod_{j=1}^{m}P_{j}(\{t_i\})}\right)\vert_{t_i=\tilde{t}_i} ={H(\{\tilde{t}_i\})_{q_1,\cdots,q_{\cal I}}\over J(\{\tilde{t}_i\})}
\ee
otherwise, a local residue can be defined for the modified integrand,
\be\label{degresidue}
{\rm Res}_{\{P_{q_1},\cdots P_{q_{\cal I}}\}}\left({G(\{t_i\}) J(\{\tilde{t}_i\})\over \prod_{j=1}^{m}P_{j}(\{t_i\})}\right)\vert_{t_i=\tilde{t}_i} =\mu_{{\{P_{q_1},\cdots P_{q_{\cal I}}\}}}(\{\tilde{t}_i\})\,H(\{\tilde{t}_i\})_{q_1,\cdots,q_{\cal I}}\,
\ee
where $\mu_{{\{P_{q_1},\cdots P_{q_{\cal I}}\}}}(\{\tilde{t}_i\})$ denotes the intersection multiplicity of $\{P_{q_1},\cdots P_{q_{\cal I}}\}$ at $\{\tilde{t}_i\}$.

By B\'ezout's theorem, the maximum number of common zeros $n_{0}$ for a system of polynomials $\{P_{q_1},\cdots P_{q_{\cal I}}\}$ is given by the product of  their degrees $n_{0}=\prod_{j=q_1}^{q_{\cal I}}{\rm deg}(P_j)$, where ${\rm deg}(P_j)$ denotes the degree of the given polynomial $P_j$. 
In order to find them, we can use the method recently developed in \cite{Cardona:2015ouc, Dolan:2015iln,Cardona:2015eba} to find solutions to the scattering equations. This method is based in both, the theory of resultants (see for example \cite{Dickenstein,Gelfand}) and the theory of elimination (see for example \cite{Sturmfels:2002}). 

In theory, the process described in this section is adequate to solve the integrals of the type \eqref{remaining}, but in practice, the application of this procedure to the actual integrals we need, namely for example \eqref{trianglemellin} and \eqref{mellinbox}, can still be quite challenging. For example, the exponents $\delta_{ij}$  correspond to the Mellin variables, or in other words, to the kinematics variables of the Mellin amplitude, and therefore take arbitrary values, so in principle we must be able to apply it to arbitrary values of the exponents. However, for the particular integrals we need to consider, some particularities might simplify or facilitate the procedure, as we shown in the appendix C,  when applied to the triangle four points loop the integral correspond to a simpler, but still involved problem.

\section{Conclusions and outlook}
In this note we have considered one loop Witten diagrams for scattering of scalars in AdS. We start by considering a contour representation for the bulk-to-bulk propagator that allow us to consider bulk and boundary coordinates in almost the same footing from the point of view of the integrated loop. By using Schwinger parametrization to exponentiate the propagators, the integration over AdS  ``radial" coordinate becomes Gaussian and can be straightforwardly integrated. With the help of a Mellin-Barnes transformation the remaining dependence on AdS coordinates can be also treated very easily.  We then Mellin transform the integral over the Schwinger parameters associated to the bulk-to-boundary propagators by means of the Symanzik formula, to extract the Mellin representation of the Witten diagram under consideration. However,  a leftover integration over the Schwinger parameters associated to the bulk-to-bulk propagators remains, and we need to look for further techniques to integrate them. We start a modest study of this leftover integrals by reinterpreting them in terms of integrals over contours  defined by the zero locus of a system of polynomial equations, which boils down to the computation of residues on a multidimensional space. 

Along the way, we realized that the reinterpretation of those Schwinger integrals in terms of contour  resembles modern formulas for the computation of scattering amplitudes in flat space in terms of contours defined by the intersection of algebraic curves in projective space, such as the Amplituheadron program \cite{ArkaniHamed:2012nw} or the most recent CHY formulation for arbitrary dimensions \cite{Cachazo:2013hca}. It would be very interesting to explore those similarities further and maybe apply this well understood techniques for scattering amplitudes in flat space to the scattering of particles in AdS spaces.

\acknowledgments
It is my pleasure to thank to Thiago Araujo for collaboration during very early stages of this work, as well as to Yu-tin Huang, Simon Caron-Huot, Ellis Ye Yuan and Yoji Koyama for useful and enlightening discussions. The work is supported in part by the Danish National Research Foundation (DNRF91) and the National Center for Theoretical Science (NCTS), Taiwan, Republic of China. 

\appendix
\section{Exchanged scalar by the split representation}
In this section we want to compute again the exchange tree-level scalar Witten diagram as an example to illustrate how the split representation works as well as to point out why this method becomes harder than the one used in this paper when it comes to the computation of loops.
The split representation have been introduced in \cite{Costa:2014kfa} and states essentially the fact that the  bulk-to-bulk propagator can be rewritten in terms of bulk-to-boundary propagators as an integration over bulk-to-boundary propagators,
\be 
\label{split}
G^{\Delta}_{B B}(X, Y)  = \int_{-i\infty}^{+i\infty} \frac{\dd c}{2\pi i}\frac{\Gamma(h+c)\Gamma(h-c) f^{\Delta}(c) }{{\cal C}_{h+c} {\cal C}_{h-c}}\int_{\d AdS}\dd K G^{(h+c)}_{\d B}(X, K) G^{(h-c)}_{\d B}(K, Y) 
\ee
The function $f(c)$ is given by,
\be 
\label{f-fuct}
f^{\Delta}(c)= \frac{1}{2\pi^{2h}\Gamma(c)\Gamma(-c)}\frac{1}{[(\Delta-h)^2-c^2]}\; .
\ee
Plugging in the above definitions back to \eqref{exchange}, we get
\be 
\begin{split}
& {\cal A}_4  (P_1, P_2, P_3, P_4) = g^2 \int_{-i\infty}^{+i\infty} \frac{\dd c}{2\pi i}\frac{\Gamma(h+c)\Gamma(h-c) f(c) }{{\cal C}_{h+c} {\cal C}_{h-c}}\int_{\d AdS}\dd K\times\\
&  \times  \int_{AdS} \dd X G_{B\d}(P_1, X) G_{B\d}(P_2, X) G_{B\d}^{(h+c)}( X, K)  \int_{AdS} \dd Y G^{(h-c)}_{\d B}(K, Y) G_{B\d}( Y, P_3) G_{\d B}(Y, P_4)\,.
\end{split}
\ee

\begin{figure}
	\centering
	\begin{subfigure}[b]{0.3\textwidth}
  \includegraphics[width=3.0cm]{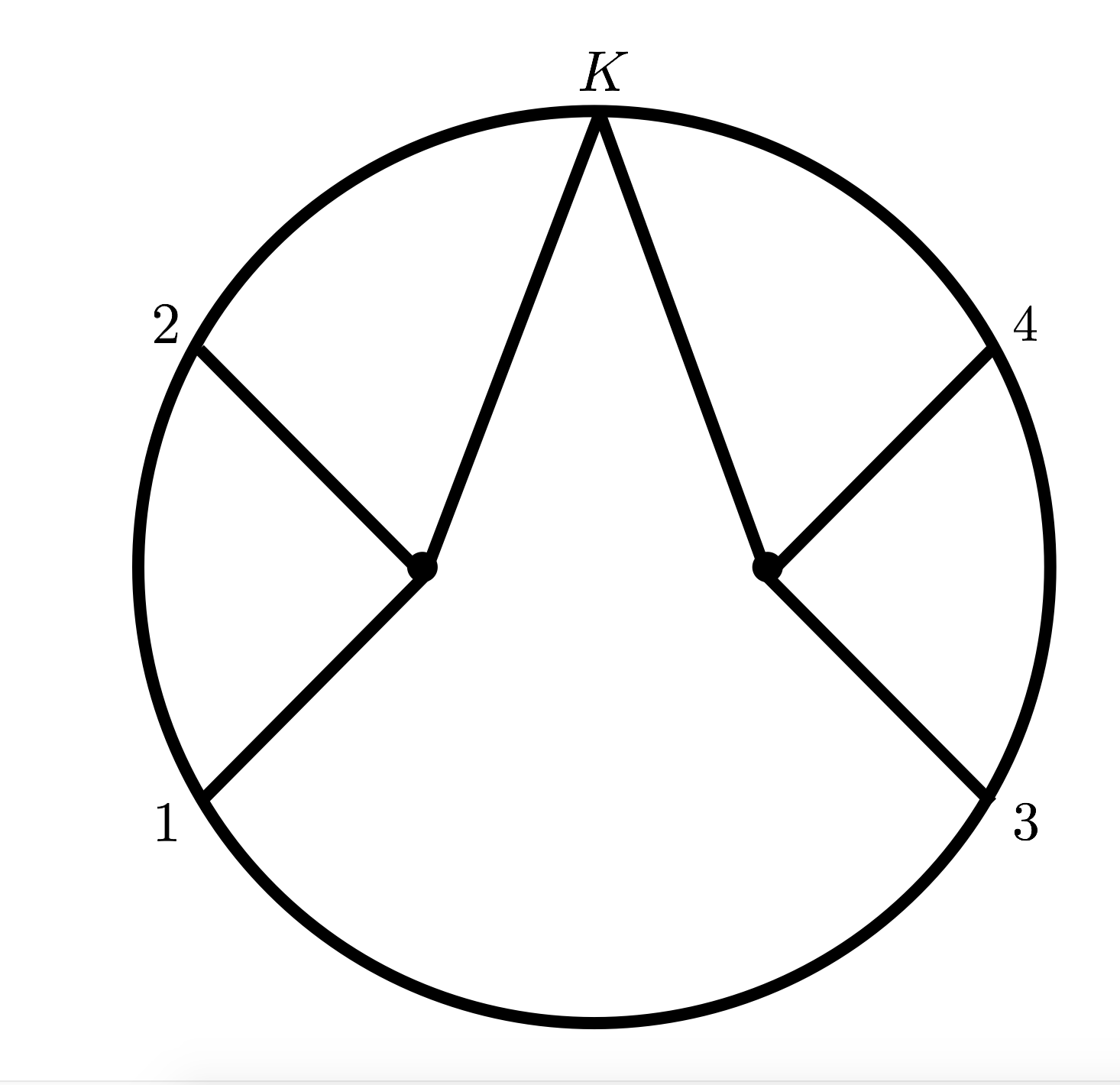}
  \caption{Scalar Exchange.}\label{exchan}
  \end{subfigure}
 	\begin{subfigure}[b]{0.3\textwidth}
  \includegraphics[width=4.0cm]{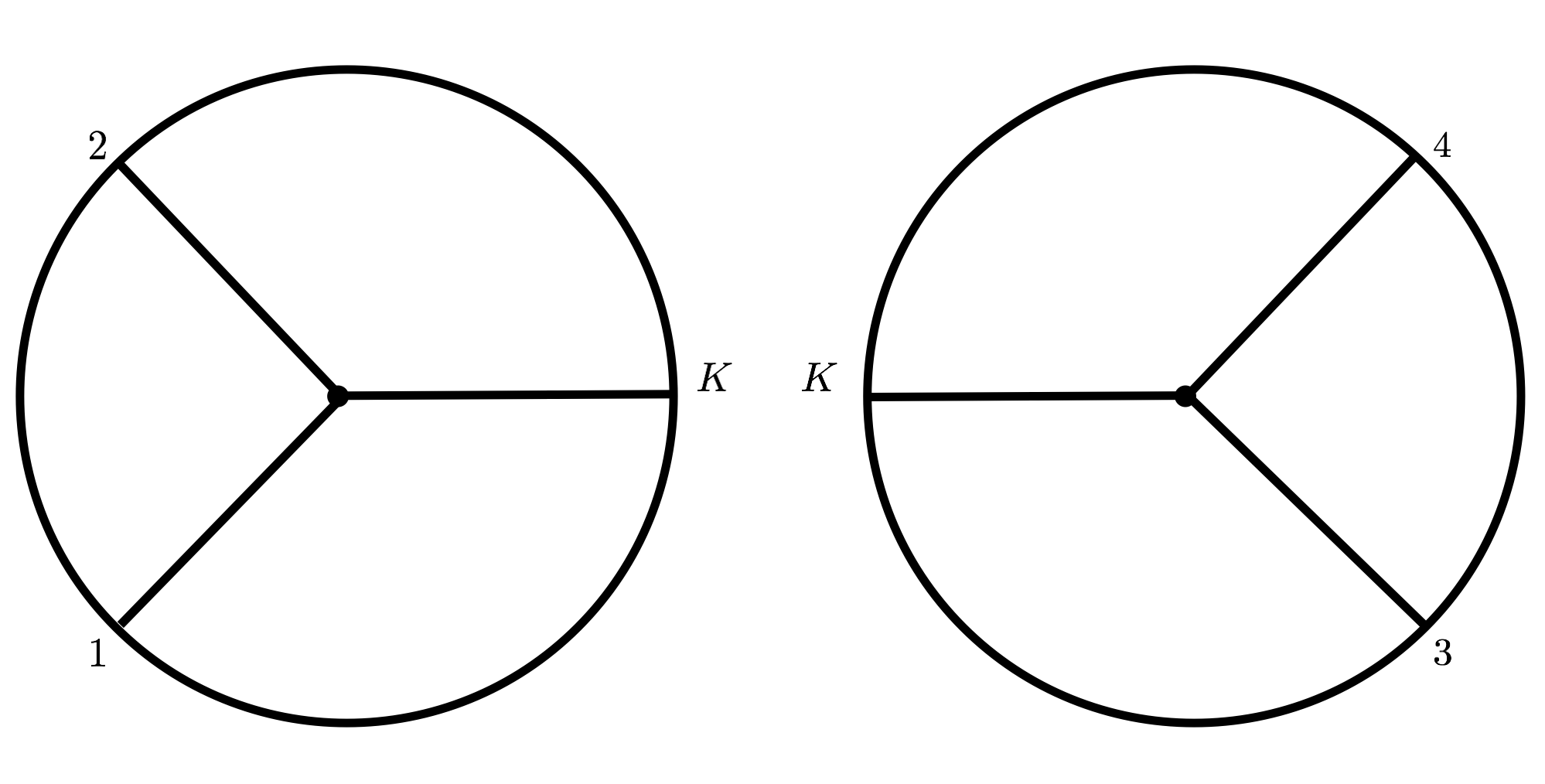}
  \caption{Product of two 3pt functions.}\label{threepoint}
  \end{subfigure} 
	\caption{Decomposition of the scalar exchange diagram.}
\end{figure}

The 4pt-amplitude is decomposed into a product of two 3pt-Witten diagrams as in Figures (\ref{exchan}) and (\ref{threepoint}) integrated over the new inserted boundary point $K$,
\be 
\label{4pf}
{\cal A}_4  (P_i) = \int_{-i\infty}^{+i\infty} \frac{\dd c}{2\pi i}\frac{\Gamma(h+c)\Gamma(h-c) f(c) }{{\cal C}_{h+c} {\cal C}_{h-c}}\int_{\d AdS}\dd K {\cal A}^{(h+c)}_3 (P_1, P_2, K) {\cal A}^{(h-c)}_3 (K, P_3, P_4) \; .
\ee
At this point is already easy to see that the splitting have trade a tree-level diagram by loop-like diagram since we have to integrate now over space-time coordinates.

The contact three-point function of Figure (\ref{threepoint}) 
\be 
\label{3pt-cont}
{\cal A}_3 (P_i, P_j, K) = g \int_{AdS}\dd X G_{B\d}(X, P_i)  G_{B\d}(X, P_j)  G_{B\d}(X, K) \; ,
\ee
is easy to computed and is given by \cite{Freedman:1998tz}
\be 
{\cal A}_3 (P_i, P_j, P_k) = g \frac{\pi^h{\cal C}_{\Delta_i} {\cal C}_{\Delta_j} {\cal C}_{\Delta_k}}{\Gamma(\Delta_i) \Gamma(\Delta_j) \Gamma(\Delta_k)}\frac{\Gamma(\Delta_{ijk}-h) \Gamma(\Delta_{ij})\Gamma(\Delta_{ik})\Gamma(\Delta_{jk}) }{P_{ij}^{\Delta_{ij}} P_{ik}^{\Delta_{ik}}  P_{jk}^{\Delta_{jk}}}
\ee

where we adopt the notation
\be 
\Delta_{ij} =\frac{\Delta_i+ \Delta_j -\Delta_k}{2}\; , \qquad  \Delta_{ijk}=\frac{\Delta_i+ \Delta_j +\Delta_k}{2}\; .\\
\ee

where $\Delta_c=h+c$ and $\Delta_{\bar{c}}=h-c$. Inserting these expressions into (\ref{4pf}), we find
\be
\label{4pf-exp} 
\begin{split} 
{\cal A}_4  (P_i) & =  \frac{g^2 \pi^{2h}}{4} \int_{-i\infty}^{+i\infty} \frac{\dd c}{2\pi i} f(c) \prod_{i=1}^4\frac{{\cal C}_{\Delta_i} }{\Gamma(\Delta_i)} \frac{1}{P_{12}^{\Delta_{12}} P_{34}^{\Delta_{34}} }
\times\\
& \times \Gamma(\Delta_{12c}-h) \Gamma(\Delta_{34\bar c}-h)\Gamma(\Delta_{12}) 
\Gamma(\Delta_{1c}) \Gamma(\Delta_{2c}) \Gamma(\Delta_{34}) \Gamma(\Delta_{3\bar c}) \Gamma(\Delta_{4\bar c})\\
& \times\int_{\d AdS}\dd K\frac{1}{(-2P_1\cdot K)^{\Delta_{1c}}} \frac{1}{(-2P_2\cdot K)^{\Delta_{2c}}} \frac{1}{(-2P_3\cdot K)^{\Delta_{3\bar c}}} \frac{1}{(-2P_4\cdot K)^{\Delta_{4\bar c}}}\; .
\end{split}
\ee
computing explicitly the boundary integral we arrive to
\be
\begin{split} 
{\cal A}_4 & (P_i)  =  \frac{g^2 \pi^{3h}}{8(2\pi i)^2} \prod_{i=1}^4\frac{{\cal C}_{\Delta_i} }{\Gamma(\Delta_i)} \int_{-i\infty}^{+i\infty} \frac{\dd c}{2\pi i} f(c) \times\\
&\times \Gamma\left(\frac{\Delta_1 + \Delta_2 + c -h }{2}\right) \Gamma\left(\frac{\Delta_3 + \Delta_4 + c -h }{2}\right) \Gamma\left(\frac{\Delta_1 + \Delta_2 - c -h }{2}\right) \Gamma\left(\frac{\Delta_3 + \Delta_4 - c -h }{2}\right)\times \\
& \times \int_{\Sigma^4}\dd \delta_{ij} \frac{\Gamma(\delta_{12}-\Delta_{12}) \Gamma(\delta_{34}-\Delta_{34})}{\Gamma(\delta_{12}) \Gamma(\delta_{34}) }
\prod_{i<j=1}^4 \Gamma(\delta_{ij})(P_{ij})^{-\delta_{ij}}\; .
\end{split}
\ee
where,
\be 
\begin{split}
&\delta_{12} = \frac{\Delta_1 + \Delta_2- s_{12}}{2}\; , \qquad\ \ \ \ \delta_{34} = \frac{\Delta_3 + \Delta_4 - s_{12}}{2}\\
& \delta_{12} -\Delta_{12} = \frac{h+c-s_{12}}{2}\; , \qquad  \delta_{34} -\Delta_{34} = \frac{h+c-s_{12}}{2}\; .
\end{split}
\ee
Therefore, 
\be
\begin{split} 
{\cal A}_4 & (P_i)  =  \frac{g^2 \pi^{3h}}{8(2\pi i)^2} \prod_{i=1}^4\frac{{\cal C}_{\Delta_i} }{\Gamma(\Delta_i)}\int_{\Sigma^4}\dd \delta_{ij} \prod_{i<j=1}^4 \Gamma(\delta_{ij})(P_{ij})^{-\delta_{ij}} \times \\
&\times \left[\Gamma\left(\frac{\Delta_1 + \Delta_2 - s_{12}}{2} \right)\Gamma\left( \frac{\Delta_3 + \Delta_4-s_{12}}{2}\right)\right]^{-1} \times \\
&\times  \int_{-i\infty}^{+i\infty} \frac{\dd c}{2\pi i} f(c)
\Gamma\left( \frac{h+c-s_{12}}{2}\right)\Gamma\left(\frac{\Delta_1 + \Delta_2 + c -h }{2}\right) \Gamma\left(\frac{\Delta_3 + \Delta_4 + c -h }{2}\right)\times\\
&  \times\Gamma\left( \frac{h-c-s_{12}}{2}\right)\Gamma\left(\frac{\Delta_1 + \Delta_2 - c -h }{2}\right) \Gamma\left(\frac{\Delta_3 + \Delta_4 - c -h }{2}\right)
\end{split}
\ee
Moreover, using that
\be 
\prod_{i=1}^4\frac{{\cal C}_{\Delta_i} }{\Gamma(\Delta_i)} = \frac{2{\cal N}}{\pi^h \Gamma\left(\frac{1}{2}\sum_{i=1}^4\Delta_i-h\right)}\; ,
\ee
and equation (\ref{f-fuct}), we find
\be
\begin{split} 
{\cal A}_4 & (P_i)  =  \frac{g^2 {\cal N}}{2(2\pi i)^2} \int_{\Sigma^4}\dd \delta_{ij} \prod_{i<j=1}^4 \Gamma(\delta_{ij})(P_{ij})^{-\delta_{ij}} \times \\
&\times \left[\Gamma\left(\frac{1}{2}\sum_{i=1}^4\Delta_i - h\right)\Gamma\left(\frac{\Delta_1 + \Delta_2 - s_{12}}{2} \right)\Gamma\left( \frac{\Delta_3 + \Delta_4-s_{12}}{2}\right)\right]^{-1} \times \\
&\times  \int_{-i\infty}^{+i\infty} \frac{\dd c}{2\pi i} \frac{1}{(\Delta-h)^2-c^2}\times \\
&\times\frac{\Gamma\left( \frac{h+c-s_{12}}{2}\right)\Gamma\left(\frac{\Delta_1 + \Delta_2 + c -h }{2}\right) \Gamma\left(\frac{\Delta_3 + \Delta_4 + c -h }{2}\right)}{2\Gamma(c)}\frac{\Gamma\left( \frac{h-c-s_{12}}{2}\right)\Gamma\left(\frac{\Delta_1 + \Delta_2 - c -h }{2}\right) \Gamma\left(\frac{\Delta_3 + \Delta_4 - c -h }{2}\right)}{2\Gamma(-c)}\; .
\end{split}
\ee
Comparing this result with the following Mellin representation
\be 
{\cal A}_4 (P_i)  =  \frac{{\cal N}}{(2\pi i)^2} \int_{\Sigma^4}\dd \delta_{ij} M(\delta_{ij}) \prod_{i<j=1}^4 \Gamma(\delta_{ij})(P_{ij})^{-\delta_{ij}} \; ,
\ee
we can see that the Mellin amplitude for the scalar exchange diagram is given by
\be 
M(\delta_{ij}) = \frac{g^2}{\Gamma\left(\frac{1}{2}\sum_{i=1}^4\Delta_i - h\right)\Gamma\left(\frac{\Delta_1 + \Delta_2 - s_{12}}{2} \right)\Gamma\left( \frac{\Delta_3 + \Delta_4-s_{12}}{2}\right)} \int_{-i\infty}^{+i\infty} \frac{\dd c}{2\pi i} \frac{\ell(c)\ell(-c)}{(\Delta-h)^2-c^2}
\ee
and 
\be 
\ell(c) \equiv \frac{\Gamma\left( \frac{h+c-s_{12}}{2}\right)\Gamma\left(\frac{\Delta_1 + \Delta_2 + c -h }{2}\right) \Gamma\left(\frac{\Delta_3 + \Delta_4 + c -h }{2}\right)}{2\Gamma(c)}\; .
\ee
which coincides with the previous result by Penedones \cite{Penedones:2010ue}.

\section{$I^{\rhd}_S$ integral in terms of Hypergeometric functions.}
In this appendix we display explicitly the integral \eqref{tvertex2} in terms of Hypergeometric functions, 
\be\label{onepagetriangle}
\begin{split}
&I^{\rhd}_S=\frac{\pi ^2 (-1)^{\gamma _{23}+f_{23}-2 l} \Gamma \left(\frac{1}{2} \left(-2 h+\Delta _1+\Delta _2+\Delta _3-1\right)\right)}{4 \Gamma \left(\Delta _1\right) \Gamma
   \left(-f_{23}-\gamma _{23}\right) \Gamma \left(\frac{1}{2} \left(-2 h+f_{12}+f_{23}+\gamma _{12}+\gamma _{23}+\Delta _2\right)\right) }\\
   &
  \frac{ \csc \left(\pi  \left(\gamma _{23}-\Delta _2+f_{23}+2 l\right)\right) \Gamma \left(\frac{1}{2} \left(2 l+f_{13}+f_{23}+\gamma _{13}+\gamma _{23}-\Delta _1-\Delta
   _2+1\right)\right) }{\Gamma \left(\frac{1}{2}
   \left(-2 h+2 l+f_{23}+f_{31}+\gamma _{23}+\gamma _{31}+\Delta _3\right)\right)}\\
   &\left(\frac{2 e^{i \pi  (3 h+l)} \Gamma \left(-h+l+\frac{1}{2}\right) \left(e^{2 i \pi  \left(\gamma _{23}+f_{23}+2 l\right)}-e^{2 i \pi  \Delta
   _2}\right)}{\left(e^{2 i \pi  \Delta _2}+e^{2 i \pi  (h+l)}\right) \left(e^{2 i \pi 
   \left(\gamma _{23}+f_{23}+l\right)}+e^{2 i \pi  h}\right)}  \Gamma \left(-h-l+\Delta _1+\Delta _2-\frac{1}{2}\right)\right.\\
   &\left.\Gamma \left(\frac{1}{2} \left(-2 h+f_{12}+f_{23}+\gamma _{12}+\gamma _{23}+\Delta
   _2\right)\right) \, _3\tilde{F}_2\left(a_1,b_1,c_1;d_1,e_1;1\right)\right.\\
   &\left.  -\Gamma \left(-f_{23}-\gamma _{23}\right) e^{i \pi  \left(\gamma _{23}+f_{23}+2 l\right)} \sec \left(\pi 
   \left(-\gamma _{23}-f_{23}+h-l\right)\right) \right.\\
   &\left.\Gamma \left(\frac{1}{2} \left(-2 l+f_{12}-f_{23}+\gamma _{12}-\gamma _{23}+\Delta _2-1\right)\right)\right.\\
   &\left. \Gamma \left(-2
   l-f_{23}-\gamma _{23}+\Delta _1+\Delta _2-1\right) \, _3\tilde{F}_2\left(a_2,b_2,c_2;d_2,e_2;1\right)\right.\\
   &\left.+e^{i \pi  \Delta _2}
   \Gamma \left(\Delta _1\right) \Gamma \left(2 l-\Delta _2+1\right) \sec \left(\pi  \left(-\Delta _2+h+l\right)\right) \right.\\   
   &\left.\Gamma \left(\frac{1}{2} \left(2
   l+f_{12}+f_{23}+\gamma _{12}+\gamma _{23}-\Delta _2+1\right)\right) \, _3\tilde{F}_2\left(a_3,b_3,c_3;d_3,e_3;1\right)\right)
\end{split}
\ee
where the tilde on the Hypergeometric functions means they are actually Hypergeometric regularized and their arguments  are given by,
\be
\begin{split}
a_1&=-h+l+\frac{1}{2},\,~b_1=\frac{1}{2} \left(-2 h+f_{12}+f_{23}+\gamma _{12}+\gamma _{23}+\Delta _2\right),\\
   c_1&=-h-l+\Delta _1+\Delta
   _2-\frac{1}{2},\,~
   d_1=-h+l+f_{23}+\gamma _{23}+\frac{3}{2},\,~
   e_1=-h-l+\Delta _2+\frac{1}{2}
\end{split} 
\ee
\be
\begin{split}
a_2&=-f_{23}-\gamma _{23},\,~b_2=\frac{1}{2} \left(-2 l+f_{12}-f_{23}+\gamma _{12}-\gamma _{23}+\Delta
   _2-1\right),\\
   c_2&=-2 l-f_{23}-\gamma _{23}+\Delta _1+\Delta _2-1,\,~
   d_2=h-l-f_{23}-\gamma _{23}+\frac{1}{2},\\
   e_2&=-2 l-f_{23}-\gamma _{23}+\Delta _2
\end{split} 
\ee
\be
\begin{split}
a_3&=\Delta _1,\,~b_3=2 l-\Delta _2+1,\,~
   c_3=\frac{1}{2} \left(2 l+f_{12}+f_{23}+\gamma
   _{12}+\gamma _{23}-\Delta _2+1\right),\\
   d_3&=h+l-\Delta _2+\frac{3}{2},\,~
   e_3=2 l+f_{23}+\gamma _{23}-\Delta _2+2\,.
\end{split} 
\ee
 \section{Few observations for the one loop triangle four-point scattering}
 Now we would like to highlight some  particularities of the triangle four points scattering.
 
Let us consider the integral at the last line of equation \eqref{trianglemellin} which we copy here for quick reference,
 \be
\begin{split}\label{triangle3}
&\int^{\infty}_{0}\prod_{i=5}^7{\dd t_i\over t_i}{1\over \Gamma(\Delta_i)}\,t_5^{g+\gamma _{13}+f_{13}-\Delta _2+\Delta_3+\Delta_4}\,t_6^{\gamma _{23}+f_{23}}\,t_7^{-\gamma _{23}-f_{23}-\Delta _3-\Delta _4-g+2 h}(1+(t_5 t_7+t_6)^2)^g\\
  &~~~~~~\left({\over}T_3(\{t_i\})\right)^{-\delta_{12}}\left({\over}T_2(\{t_i\})\right)^{-\delta_{13}-\delta_{14}}\left({\over}T_1(\{t_i\})\right)^{-\delta_{23}-\delta_{24}}\,,
  \end{split}
\ee
where as before,
\be\label{titriangle}
\begin{split}
T_1(\{t_i\})&=t_5 t_7+t_6 ,\,~~~T_2(\{t_i\})=t_7+t_5(t_5 t_7+t_6),\\
T_3(\{t_i\})&=t_5+(t_7+t_5(t_5 t_7+t_6))(t_5 t_7+t_6)
\end{split}
\ee
Since in this case the number of polynomials $T_j$ equals the number of integration variables, we pick them to compute the Jacobian
which quite interestingly is not only non-degenerate but even more, equals to one!.
That allow us to perform a change of variables $x_i=T_i$, to rewrite the integral as,
 \be
\begin{split}
&\int^{\infty}_{0}\prod_{i=5}^7{\dd x_i\over \Gamma(\Delta_i)}(x_5^2+1)^{g-1} (x_5^3 x_6^2-2 x_5^2 x_6 x_7+x_5 (x_6^2+x_7^2+1)-x_6
   x_7)^{\gamma_{23}+f_{23}-1}\\
   &~~~~~~~~~ (x_7-x_5 x_6)^{\gamma_{13}-\Delta_2+\Delta_3+\Delta_4+f_{13}+g-1}\\
  &~~~~~~{1\over x_7^{\delta_{12}}\, x_6^{\delta_{13}+\delta_{14}}x_5^{\delta_{23}+\delta_{24}}\,(x_5(x_5 x_6-
   x_7)+x_6)^{\gamma_{23}+\Delta_3+\Delta_4+f_{23}+g-2 h+1}}\,,
  \end{split}
\ee
from here is clear that the only poles comes from the simpler equations $x_i=0,\,~i=5,6,7.$ Deforming the contour such as it enclosed the poles around $x_i=0,\,~i=5,6,7,$ we have,
 \be\label{trianglelast}
\int_{{\cal C}}\,\prod_{i=5}^7{\dd x_i\over \Gamma(\Delta_i)}{G(x_5,x_6,x_7)\over x_7^{\delta_{12}}\, x_6^{\delta_{13}+\delta_{14}}x_5^{\delta_{23}+\delta_{24}}\,(x_5(x_5 x_6-
   x_7)+x_6)^{\gamma_{23}+\Delta_3+\Delta_4+f_{23}+g-2 h+1}}\,,
\ee
where we have defined 
\be
\begin{split}
G(x_5,x_6,x_7)&=(x_5^2+1)^{g-1} (x_5^2x_6(x_5 x_6-2  x_7)+x_5 (x_6^2+x_7^2+1)-x_6
   x_7)^{\gamma_{23}+f_{23}-1} \\
   &~~~~(x_7-x_5 x_6)^{\gamma_{13}-\Delta_2+\Delta_3+\Delta_4+f_{13}+g-1}\,.
   \end{split}
\ee
Even thought the integration \eqref{trianglelast} is simpler that \eqref{triangle3}, on the surfaces satisfying 
\be J(0)={\rm Det}\left({\partial (P_{q_1},\,P_{q_{2}}, P_{q_{3}})\over \partial(\{x_i\})}\right)(0)=0\,,\ee 
where
\be
 (P_{q_1},\,P_{q_{2}}, P_{q_{3}})\in \{x_7^{\delta_{12}},\, x_6^{\delta_{13}+\delta_{14}},\,x_5^{\delta_{23}+\delta_{24}},\,(x_5(x_5 x_6-
   x_7)+x_6)^{\gamma_{23}+\Delta_3+\Delta_4+f_{23}+g-2 h+1}\}\,,
\ee
i.e, on the surfaces where the Jacobians degenerates, we can not write a simple formula as \eqref{nondegresidue}. However, we can still used iteratively the Cauchy integral formula on each $x_i$, which for a given value of $\delta_{ij}$ will give us a sum of rational functions of gamma functions.  
\bibliographystyle{JHEP}
\bibliography{LMSE}
\end{document}